\documentclass[12pt]{iopart}
\usepackage{graphics,epsfig,subfig}
\usepackage{amssymb,amsfonts}
\usepackage{color,psfrag}
\usepackage{iopams}

\newcommand{\bin}[2]{{#1 \choose #2}}
\newcommand{\nn}{\nonumber}
\newcommand{\bb}{\begin{eqnarray}}
\newcommand{\ee}{\end{eqnarray}}

\newcommand{\efig}[1]{Fig. \ref{#1}}
\renewcommand{\eref}[1]{Eq. (\ref{#1})}

\newcommand{\ff}{\frac{1}{2}}

\newcommand{\EE}{x}
\newcommand{\EF}{y}
\newcommand{\drm}{{\rm d}}
\renewcommand{\fl}{\hspace{-1cm}}

\newcommand{\xn}{\alpha_n}

\newcommand{\de}{\tilde\rho}

\newcommand{\pol}{{\cal Q}}
\newcommand{\Px}{P_x}
\newcommand{\com}{\textcolor{black}}

\begin{document}
\title{Modified stochastic fragmentation of an interval as an aging process}
\author{Jean-Yves Fortin}
\address
{\com{Laboratoire de Physique et Chimie Th\'eoriques,
CNRS UMR 7019, 
\\ Universit\'e de Lorraine,
54506 Vandoeuvre-l\`es-Nancy, France}
}
\ead{jean-yves.fortin@univ-lorraine.fr}
\begin{abstract}{
\com{
We study a stochastic model based on a modified fragmentation of a finite interval. The mechanism consists in cutting the interval at a random location and substituting a unique fragment on the right of the cut to regenerate and preserve the interval length. This leads to a set of segments of random sizes, with the accumulation of small fragments near the origin. This model is an example of record dynamics, with the presence of "quakes" and slow dynamics. The fragment size distribution is a universal inverse power law with logarithmic corrections. The exact distribution for the fragment number as function of time is simply related to the unsigned Stirling numbers of the first kind. Two-time correlation functions are defined and computed exactly. They satisfy scaling relations and exhibit aging phenomena. In particular the probability that the same number of fragments is found at two different times $t>s$ is 
asymptotically equal to $[4\pi\log(s)]^{-1/2}$ when $s\gg 1$ and the ratio $t/s$ 
fixed, in agreement with the numerical simulations. The same process with a
reset impedes the aging phenomena beyond a typical time scale defined by the reset parameter.}}
\end{abstract}

\pacs{05.40.-a, 64.60.av, 64.60.Ht}

\maketitle
%
%
\section{Introduction}
Stochastic models present various dynamical properties that can be classified according
to the long time behavior of their physical observables such as order parameters or correlation functions. 
For aging models, the state of the system at a given time depends strongly on
how initially it is quenched (by temperature or magnetic field for example), and presents a slow and non-exponential relaxation behavior seen in the correlation functions, and an absence of time-translation invariance \cite{struick77,book:henkel_aging,book:henkel_noneq,crisanti03}. Slow dynamics  and relaxation towards equilibrium are characteristic of glassy systems, which can be observed in spin glasses, polymers, gels, and display typical out of equilibrium properties \cite{book:Struick}. 
On the other hand, a fragmentation process is, unless it is stopped at some point, also 
classified as an out of equilibrium process, often characterized by scale-invariance of the fragment sizes \cite{krapivsky00}, which leads to numerous studies and applications in mass fragmentation in astronomy, aerosolization, and geology \cite{book:turcotte,brown89,herrmann06,bershadskii00}. Non-Gaussian stationary distributions of the fragment sizes for example are inherent to the fragmentation process, for example log-normal, Weibull or stretched exponentials, and power-law distributions
are often found numerically or empirically. These distributions 
depend on the fragmentation process itself \cite{ziff:86,hassan:95,jyf:2013}. 

Modified fragmentation is provided by different mechanisms \cite{book:krapivsky,book:bertoin}, ranging from coagulation, aggregation, to reversible polymerisation, or recombination/shuffling in DNA sequences \cite{stemmer94,cohen10}. A stochastic fragmentation can also be followed by a growth process or recombination. This phenomena can be found in biology where for example a segmented planarian organism  regenerates its body \cite{lobo12}. Fragmentation/growth mechanism is also studied like here in interval models that display absorbing states and where the distribution of the times of absorption \cite{azais15} is a quantity of interest. In chemistry, kinetics of aggregate fragmentation is combined with reversible aggregation \cite{ke03,ke04}. The aggregate size distribution in this case possesses scaling properties which depend on the reaction rates.

\com{In the following section \ref{sec_master}, we introduce a model of fragmentation with an additional mechanism which mimics a growth or recombination phenomena, in the sense that the segments on the right side of the fracture are discarded and replaced by a new and unique segment that regenerates the interval length, preventing the total fragmentation of the interval. 
More importantly this asymmetric process induces a slow dynamics in the sense that the evolution of the observable, here the average number of fragments, is governed by a growth law which is logarithmic with time, as well as the variance. 
This is due to the presence of
rare events or "quakes" that slow down the multiplication of fragments. Normally the system produces more and more fragments, but rare events (records) send the system back to a two fragments configuration, however with a probability decreasing with time because more and more small fragments are produced near the origin of the interval. A state with a two fragments configuration requires therefore that the fragmentation has to be located more and more on the far left with a decreasing probability in time, as we shall see.
\break
The physics of this model presents similarity with models of aging based on record dynamics theory \cite{robe:16}, which attempts to explain the jamming phenomena or irreversible domain rearrangements in supercooled colloid liquids near the glass transition \cite{yunker:09}. One
description of non-equilibrium dynamics in glassy systems involves log-Poisson statistics for the number of records, or their time distribution\cite{sibani:93,sibani:03,sibani:16}. These records are principally driven by noise adaptation that allows the system to pass from one free energy valley to another. The model of fragmentation we propose in this paper does not possess any energy landscape, because there is no Hamiltonian, but still involves glassy phenomena. Glassy behavior without energy barriers exists in theoretical models such as the backgammon model \cite{ritort:95,franz:96,crisanti03} for which the relaxation towards the ground state (all the boxes empty except one) is slowed down by the fact that it is harder to empty the remaining boxes for which the particle density increases with time, as the total number of particles is fixed. The relaxation time in this model is governed by entropic barriers, which lead to critical slowing down if the available states of lower energy are decreasing with time. This is for example the case of the glassy transition seen in colloidal hard spheres \cite{megen:94}. Another example leading to a glassy behavior is provided by the East model \cite{jackle:91,faggionato:12}. Ising spins one a one-dimensional chain are allowed to flip if their neighbor on the right (east) is in the up position. The asymmetric constraint in the chain is the cause of the slow dynamics observed in this system, and the auto-correlation of the first spin is characterized by a stretched exponential. The asymmetry property is also characteristic of our model, with the slow accumulation near the origin of small fragments which have more the chance to survive.}

As we will demonstrate in section \ref{sec_frag_numb}, the master equation of this model can be solved exactly, and 
the average number of fragments is expressed simply as function of the unsigned 
Stirling numbers of first kind. The fragment size distribution computed in section \ref{sec_frag_size} follows asymptotically an inverse power law. The main part of this paper is described in section \ref{sec_correlation}, where we introduce two-time correlation functions, given by the probability to have a configuration 
of equal fragment number at two different times. This quantity presents aging properties typical of glassy systems, and can be computed exactly. In the long time limit, the leading terms are sufficient to explain the numerical simulations performed for this model.
Finally, in section \ref{sec_reset}, a reset mechanism is introduced and we explicitly show that the aging phenomena is replaced by a regime of fast relaxation.

%
\section{Master equation and solution\label{sec_master}}
\begin{figure}
\centering
\subfloat[ ]
{
\includegraphics[scale=1.2,clip]{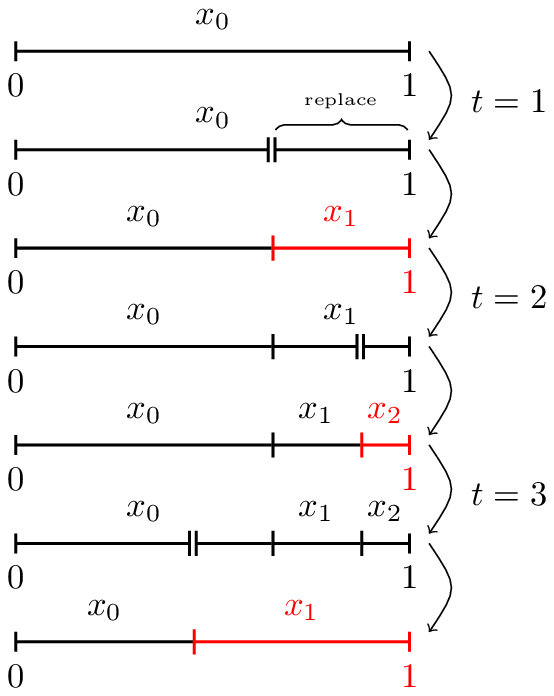}
\label{fig_process}
}
\subfloat[ ]
{
\includegraphics[scale=1.2,clip]{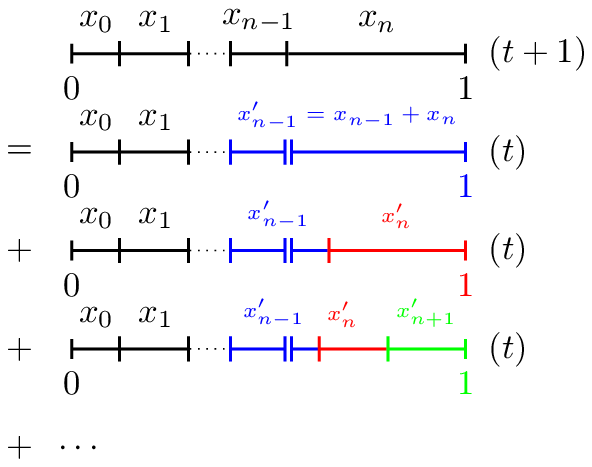}
\label{fig_master}
}
\caption{(a) Process of fragmentation/substitution on an interval initially of 
size $\EE_0$. After one iteration, the interval is fragmented into two pieces (double vertical bar), 
at a random location. All fragments on the right part are removed and replaced by a new and unique 
fragment. (b) Diagrammatic representation of the master equation at time $t+1$ 
in terms of processes occurring at previous time $t$.}
\end{figure}

\begin{figure}
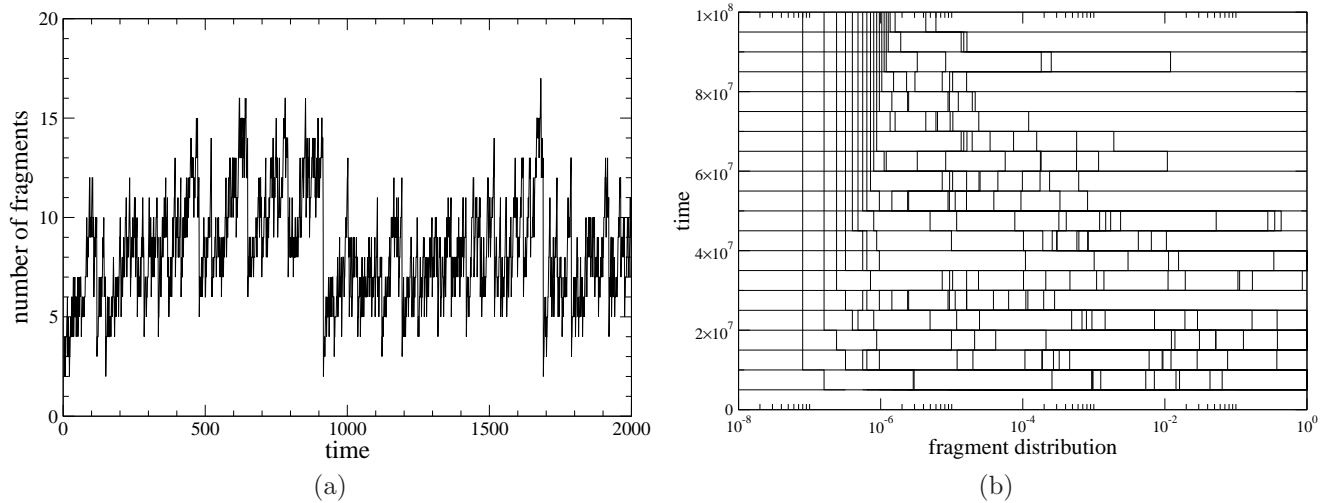

\centering
\subfloat[ ]
{
\includegraphics[scale=0.35,clip]{fig_levels_bis}
\label{fig_levels_bis}
}
\subfloat[ ]
{
\includegraphics[scale=0.35,clip]{fig_frag}
\label{fig_frag}
}
\caption{(a) Evolution of the random fragmentation process showing the number of $(n+1)$ fragments as function of time up to 2000 iterations. Rare events or "quakes" occur when the number of fragments, after increasing, is reduced to only 2.\com{ (b) Typical fragment configurations on the interval, for times up to $10^8$ iterations. Small fragments accumulate near the origin.}}
\end{figure}

%
We consider an interval of size unity, initially filled with only one fragment 
of size $\EE_0=1$. The stochastic process studied in the following consists in choosing a
random number $x$ between 0 and 1 within a uniform
distribution, at every time increment, and in removing the right part $[x,1]$ on the interval. This part 
is then replaced by a new fragment of size $1-x$. This is exemplified in \efig{fig_process}, 
where the process is described for three time steps. At time $t=1$, the initial 
fragment $\EE_0=1$ is cut into two parts, such that the right part is replaced 
by a new fragment of size $\EE_1$ with $\EE_0+\EE_1=1$. Then a new random 
number $x$ is chosen, whose value $\EE_0<x<1$ falls into the second 
fragment of size $\EE_1$. The right part discarded and replaced by a new 
fragment of size $\EE_2$, with now $\EE_0+\EE_1+\EE_2=1$. Repeating the process 
at $t=3$ with a random number falling into the first fragment $\EE_0$ leads to a
new fragment $\EE_1$, with $\EE_0+\EE_1=1$. 
To quantify the dynamics, we introduce the probability distribution
$P_t^{(n)}(\EE_0,\EE_1,\cdots,\EE_n)$ of having at time $t$ exactly
$n+1$ fragments of size $\EE_0$,$\cdots$,$\EE_n$, with the constraint
$\EE_0+\cdots+\EE_n=1$. This probability satisfies a discrete time equation 
given by a sum of previous contributions
\bb\nn\fl
P^{(n)}_{t+1}(\EE_0,\cdots,\EE_n)=P^{(n-1)}_{t}(\EE_0,\cdots,\EE_{n-2},\EE_{n-1}
+\EE_n)
\\ \nn\fl
+\int_{\EE_{n-1}}^{\EE_{n-1}+\EE_n}\drm\EE'_{n-1}P^{(n)}_{t}(\EE_0,\cdots,\EE_{n-2
} , \EE'_{n-1}, \EE_{n-1}+\EE_n-\EE'_{n-1})
\\ \nn
\fl\fl
+\int_{\EE_{n-1}}^{\EE_{n-1}+\EE_n}\drm\EE'_{n-1}
\int_{0}^{\EE_{n-1}+\EE_n-\EE'_{n-1}}\drm\EE'_{n}
P^{(n+1)}_{t}(\EE_0,\cdots,\EE_{n-2},\EE'_{n-1},\EE'_{n},\EE_{n-1}+\EE_n-
\EE'_{n-1}-\EE'_{n})
\\ \fl\label{eq_master}
+\cdots
\ee
The meaning of the different contributions is explained in \efig{fig_master}. 
The configuration $(x_0,\cdots,x_n)$ of $(n+1)$ segments at time $t+1$
originates from different possible configurations at previous time $t$. The first 
term on the right hand side of \eref{eq_master} and \efig{fig_master} is 
one contribution from a configuration of $n$ segments whose last
segment $x'_{n-1}$ was cut into two parts. The second term comes from a 
configuration of $n+1$ segments whose $n^{th}$ segment one was also cut into two parts,
and the right part of the cut was discarded and replaced by a unique segment of size $x_n$.
In general, a configuration of $(n+1)$ segments $(x_0,\cdots,x_n)$ at time $t+1$ can originate from different 
configurations at time $t$ that had an arbitrary number of segments on the right
hand side of the $n^{th}$ segment. This process is therefore slow since the number of segments can
not grow uniformly, as seen in \efig{fig_levels_bis}. Small fragments accumulate with time near the origin as displayed in \efig{fig_frag}, which implies it is more and more difficult to find a configuration with only two fragments since the fracture has to take place at the leftmost side of the interval.  
Also, at most $t+1$ segments or elements can be produced at the given
time $t$. From the initial condition $P_0^{(0)}(x_0)=\delta(x_0-1)$ and 
\eref{eq_master}, we can deduce recursively all probability 
functions. For example, one 
obtains $P_1^{(1)}(x_0,x_1)=P_0^{(0)}(x_0+x_1)=\delta(x_0+x_1-1)$.
In order to solve \eref{eq_master}, it is convenient in the following to consider the sum variables
\bb
u_k=\sum_{i=k}^nx_i,\; u_0=x_0+\cdots+x_n=1
\ee
These variables are ordered, since $1=u_0\ge 
u_1\ge\cdots u_{n-1}\ge u_n\ge 0$. Therefore we can rewrite the distribution as
$P_t^{(n)}(\EE_0,\cdots,\EE_n)=\delta(u_0-1)Q_t^{(n)}(u_1,\cdots,u_n)$, with

\bb\nn
Q^{(n)}_{t+1}(u_1,\cdots,u_n)=Q^{(n-1)}_{t}(u_1,\cdots,u_{n-1})
\\ \label{eq_master2}
+\sum_{k=1}^{t-n+1}\int_{0}^{u_n}\drm z_1\int_0^{z_1}\drm z_2\cdots
\int_0^{z_{k-1}}\drm z_{k}
Q^{(n+k-1)}_{t}(u_1,\cdots,u_{n-1},z_1,\cdots,z_k).
\ee
%
%

%
The general solution, starting with initial condition $Q_0^{(0)}=1$, is given by
\bb\nn
Q_t^{(n)}(u_1,\cdots,u_n)=
\sum_{k_1+k_2+\cdots+k_n=t-n}u_1^{k_1}\cdots u_n^{k_n}
\\ \label{Q_gen}
=
\sum_{k_1=0}^{t-n}u_1^{
t-n-k_1}\sum_{k_2=0}^{k_1}u_2^{k_1-k_2}\cdots
\sum_{k_{n-2}=0}^{k_{n-3}}u_{n-2}^{k_{n-3}-k_{n-2}}
\sum_{k_{n-1}=0}^{k_{n-2}}u_{n-1}^{k_{n-2}-k_{n-1}}u_n^{k_{n-1}}.
\ee
This polynomial solution can be deduced by induction from \eref{eq_master2}. 
Indeed, the first terms are given explicitly by 
\bb\nn
Q_1^{(1)}(u_1)=1,
\\ \nn
Q_2^{(1)}(u_1)=u_1,\;Q_2^{(2)}(u_1,u_2)=1,
\\ \nn
Q_3^{(1)}(u_1)=u_1^2,\;Q_3^{(2)}(u_1,u_2)=u_1+u_2,\;Q_3^{(3)}(u_1,u_2,u_3)=1,
\\ \nn
Q_4^{(1)}(u_1)=u_1^3,\;Q_4^{(2)}(u_1,u_2)=u_1^2+u_1u_2+u_2^2,
\\ \nn
Q_4^{(3)}(u_1,u_2,u_3)=u_1+u_2+u_3,\;Q_4^{(4)}(u_1,u_2,u_3,u_4)=1,
\\ \nn
Q_5^{(1)}(u_1)=u_1^4,\;Q_5^{(2)}(u_1,u_2)=u_1^3+u_1^2u_2+u_1u_2^2+u_2^3,
\\ \nn
Q_5^{(3)}(u_1,u_2,u_3)=u_1^2+u_2^2+u_3^2+u_1u_2+u_1u_3+u_2u_3,
\\ \nn
Q_5^{(4)}(u_1,u_2,u_3,u_4)=u_1+u_2+u_3+u_4,
\\ \label{Q15}
Q_5^{(5)}(u_1,u_2,u_3,u_4,u_5)=1.
\ee
From these examples, we can infer that $Q_t^{(n)}(u_1,\cdots,u_n)$ is a sum of monoms 
of order $t-n$ with coefficients unity, with all possible symmetric combinations between the variables $u_k$. In 
particular $Q_t^{(t)}=1$ is a constant equal to unity. We also have by normalization
\bb\nn
\sum_{n=1}^{t}\int_0^1\drm u_1\int_0^{u_1}\drm u_2\cdots \int_0^{u_{n-1}}
\drm u_n Q_t^{(n)}(u_1,\cdots,u_n)=1
\ee
%
\section{Fragment number distribution\label{sec_frag_numb}}
%
In this section, we are interested in the probability $\rho_n(t)$ to have $n+1$ segments (or fragments) at time $t$. This 
quantity is given by performing the multiple integral over all the $u_i$s
\bb\nn
\rho_n(t)&=&\int_0^1\drm u_1\int_0^{u_1}\drm u_2\cdots 
\int_0^{u_{n-1}}\drm u_nQ_t^{(n)}(u_1,\cdots,u_n)
\\ \label{rho_def}
&=&\frac{1}{n!}\int_0^1\drm u_1\int_0^{1}\drm u_2\cdots 
\int_0^{1}\drm u_nQ_t^{(n)}(u_1,\cdots,u_n)
\ee
with the normalization $\sum_{n=1}^{t}\rho_n(t)=1$.
The factor $n!$ in the last part comes from the redefinition of the 
integral domains, which is possible in this case since 
$Q_t^{(n)}(u_1,\cdots,u_n)$ is a symmetric polynomial by exchange of 
its variables, as explicitly shown in \eref{Q15}. Using the Kronecker delta function representation
$\int_{-\pi}^{\pi}\frac{\drm y}{2\pi}\e^{iyk}=\delta_{k,0}$, one can decouple the 
integers $k_i$ in the constraint of \eref{Q_gen}, and perform the summation from $k_i=0$ to $k_i=\infty$, then integrate over the 
$u_i$ variables. Therefore the distribution can be rewritten as
\bb\nn
\rho_n(t)=\frac{1}{n!}\int_{-\pi}^{\pi}\frac{\drm y}{2\pi}\e^{iy(t-n)}
\left (\int_0^1\drm u_1 \sum_{k_1=0}^{\infty}u_1^{k_1}\e^{-iyk_1}\right )
\cdots
\left (\int_0^1\drm u_n \sum_{k_n=0}^{\infty}u_n^{k_n}\e^{-iyk_n}\right )
\\ \label{rho_n_int}
=\frac{1}{n!}\int_{-\pi}^{\pi}\frac{\drm y}{2\pi}\e^{iyt}
\left [-\log(1-\e^{-iy})\right ]^n
\ee
We can in principle replace the upper limit of the $k_i$, $t-n$, by $\infty$, but this leads to some singular behavior at $\e^{-iy}=1$ in \eref{rho_n_int}, which can be removed using a complex contour integral $z=\e^{-iy}$ on the unit circle excluding the point $z=1$. Another possibility is to use a parameter $\rho<1$ and replace $z\rightarrow \rho z$. The 
limit $\rho=1$ is implicitly taken after all the integrations are performed. It is then useful to consider
the expansion of the logarithmic function in terms of unsigned Stirling
numbers of the first kind $s(r,n)>0$, with $r\ge n\ge 0$
\bb\label{def_stirling}
[-\log(1-z)]^n=n!\sum_{r=n}^{\infty}\frac{s(r,n)}{r!}z^r
\ee
This allows us to express simply $\rho_n(t)$, after integration over $y$, as
\bb\label{def_rho_n}
\rho_n(t)=\frac{s(t,n)}{t!},\;\rho_0(t)=\delta_{t,0}
\ee
It is zero for $n>t$ as expected. The Stirling number of the first kind $s(t,n)$ is the 
number of ways of partitioning $t$ points in $n$ oriented cycles.
The distribution $\rho_n(t)$ is normalized 
$\sum_{n\ge 1}^{t}\rho_n(t)=1$, using the Stirling number identities \cite{Abram_Stirl,Gould}. 
These numbers, can be found in models of discrete fragmentation as they appear similarly in the 
expression of the average number of fragments of a given size \cite{Delannay:1996}.
They appear also naturally in stochastic problems 
involving the number of records in a sequence of random numbers 
$x_{i=1,..,t}$, at a given time $t$ \cite{Luck:2008}. The $x_i$ are chosen independently 
and identically among a uniform distribution at every successive time step 
from time 1 to $t$.  The number of records among the $x_i$ is 
equal to the number of times $x_i$ is larger than the preceding variables 
$x_{i-1},..,x_{1}$. The probability than $n$ records occur during time $t$, given that $x_1$ is 
always a record, is equal to $s(t,n)/t!$, as \eref{def_rho_n}. The distribution of the number of fragments is therefore related to a record process. 
Another class of stochastic models in close relation with Stirling numbers concerns random geometric series \cite{BenNaim:2002,BenNaim:2004}, where a growing sequence of terms $\{x_n\}_{n=0,..,t}$ at time $t$ is constructed from a random process, for example $x_t=2x_p$, with $x_p$, $0\le p\le t-1$, one of the previous term chosen with a given probability and multiplied by a factor 2. In this case the probability distribution of the logarithmic sequence depends explicitly on Stirling numbers. 

Now we are interested in the limit distribution of
$\rho_n(t)$ when $t$ is large. There are several results for the asymptotic limit of
$s(t,n)$ \cite{Wilf:93,Temme:93}, and we use a similar approach here. Following the work by Wilf \cite{Wilf:93},
a good approximation for long times $t$ is given by
\bb\label{Wilf_eq}
\rho_n(t)=\frac{1}{t}\left [\sum_{i=1}^n\gamma_i\frac{(\log t)^{n-i}}{(n-i)!}+{\cal O}\left (
\frac{(\log t)^{n-2}}{t} \right ) \right ]
\ee
where $\gamma_i$ are the coefficients of the series expansion of the inverse gamma function
\bb\nn
\frac{1}{\Gamma(z)}=\sum_{i\ge 1}\gamma_i z^i=z+\gamma z^2+\left (\frac{\gamma^2}{2}-\frac{\pi^2}{12}
\right )z^3+\cdots
\ee
\com{The result \eref{Wilf_eq} shows that the probability of having $n+1$ fragments is governed by a sum of log-Poisson distributions for different number of events, up to $n-1$, and weighted by the $\gamma_i$ coefficients. The log-Poisson term comes from the fact that the time variable 
is replaced by $\log t$. This distribution appears in theories of aging based on record dynamics and noise adaptation \cite{sibani:93,sibani:03}. }
The expansion \eref{Wilf_eq} is very useful for efficient numerical evaluation of $\rho_n(t)$ (and subsequently correlation functions), but not a priori practical for finding the limit distribution based a rescaling of the variable $n$
around the average value. 
We use here a different approach in term of saddle point analysis in order to compute the limit distribution of $\rho_n(t)$. Indeed the first question is to determine the universal behavior of this distribution in the long time limit. This means
first to find the typical average number of fragments and their variance. 
The average number of fragments $\langle n \rangle_t$ as function of time  is given by a contour integral around the unit circle
\bb\nn
\langle n \rangle_t=\sum_{n=1}^tn\rho_n(t)=\oint\frac{\drm z}{2i\pi z^{t+1}}
\frac{[-\log(1-z)]}{1-z}
\\ \label{n_mean}
=\oint\frac{\drm z}{2i\pi z^{t+1}(1-z)}
\sum_{r=1}^{\infty}\frac{s(r,1)}{r!}
z^r=\sum_{r=1}^t\frac{1}{r}\simeq \log(t)+\gamma
\ee
since $s(r,1)=(r-1)!$. Here we have excluded $z=1$ (or $y=0$) in the contour around the unit circle as explained previously. It will be implicitly assumed in the following for all calculations involving a complex integral. The variance is
defined by $\sigma_t^2=\langle n^2\rangle_t-\langle n\rangle_t^2$, and can be expressed with the identity
%
\com{
\bb\fl \nn
\langle n(n-1)\rangle_t=\sum_{n\ge 2}n(n-1)\rho_n(t)=
\sum_{n\ge 2}\frac{1}{(n-2)!}\int_{-\pi}^{\pi}\frac{\drm y}{2\pi}\e^{iyt}
\left [-\log(1-\e^{-iy})\right ]^{n-2+2}
\\ 
=\int_{-\pi}^{\pi}\frac{\drm y}{2\pi}\frac{\e^{iyt}}{1-\e^{-iy}}
\left [-\log(1-\e^{-iy})\right ]^{2}
=2!\sum_{r\ge 2}\frac{s(r,2)}{r!}\int_{-\pi}^{\pi}\frac{\drm y}{2\pi}\frac{\e^{iy(t-r)}}{1-\e^{-iy}}
\ee
The integration over $y$ can be performed formally, using the previous prescription, by expanding simply $(1-\e^{-iy})^{-1}=\sum_{s\ge 0}\e^{-iys}$, and integrating over $y$. Each integration gives a Kronecker function $\delta_{t-r-s}$ which is non zero if $s=t-r$. This selects however only terms for which $r\le t$ since $s\ge 0$.
Therefore one obtains, after identifying $s(r,2)/r!=\rho_2(r)$}
\bb\label{srho2}
\langle n(n-1)\rangle_t=2!\sum_{r=2}^{t}\rho_2(r)
\ee
\com{
Using formula \eref{Wilf_eq}, with $\rho_2(r)\simeq [\log(r)+\gamma]/r$ for large $r$, one has the logarithmic asymptotic behavior
$\sigma_t\simeq \sqrt{\log(t)}$. To obtain the corrective constant to this dominant behavior, we would need a more precise evaluation of $\rho_2(r)$ for $r$ not too large as many terms with $r$ finite in \eref{srho2} contribute to the sum.
Instead, one considers the average value $\langle n^2\rangle_t$ in the large time limit, which is directly expressed from \eref{Wilf_eq} 
\bb\nn
\langle n^2\rangle_t\simeq \frac{1}{t}\sum_{n=1}^{\infty}n^2\sum_{i=1}^n\gamma_i
\frac{(\log t)^{n-i}}{(n-i)!}
\ee
By reorganizing this double sum \footnote{One has $\sum_{n=1}^{\infty}a_n\sum_{i=1}^n\gamma_ib_{n-i}=\sum_{n=0}^{\infty}b_n\sum_{i=1}^{\infty}\gamma_ia_{n+i}$}, we can put
this expression in a more convenient asymptotic form
\bb\nn\fl
\langle n^2\rangle_t
\simeq \frac{1}{t}\sum_{n=0}^{\infty}\frac{(\log t)^{n}}{n!}\sum_{i=1}^{\infty}\gamma_i(n+i)^2
=\log(t)[1+\log(t)]+2\gamma \log(t)+\gamma+\gamma^2-\pi^2/6
\ee
using the identities $\sum_{i\ge 1}\gamma_i=1$, $\sum_{i\ge 1}i\gamma_i=\gamma$, and $\sum_{i\ge 1}i^2\gamma_i=\gamma+\gamma^2-\pi^2/6$. Finally one obtains a more precise asymptotic behavior for the variance after subtracting $\langle n\rangle_t^2$ from the previous expression}
\bb\label{sigma_t}
\sigma_t^2\simeq \log(t)+\gamma-\pi^2/6
\ee
The average and variance of the number of fragments follow the same law as for the problem of random geometric series \cite{BenNaim:2004}. Indeed, if one takes for example the time series  $x_t=2x_p$, with $0\le p\le t-1$ chosen randomly and initial condition $x_0=1$, considering instead the variable $n_t=\log(x_t)/\log(2)$ leads to the recurrence $n_t=n_p+1$. It is found that the average and variance of $n_t$ are the same as the equations above \eref{n_mean} and \eref{sigma_t}. The reason is that the recurrence for $n_t$ should be equal to the recurrence for the number of fragments at a given time $t$, since the number of fragments can grow by one unit at most, or equal to one of the previous generation of fragment numbers and chosen (uniformly) randomly.
Using these results, an asymptotic estimate of \eref{def_rho_n} or Stirling numbers can be done, based on a saddle point analysis
which was performed by Temme \cite{Temme:93}, but here we will use a different
form for the argument function to extremize. We would like indeed an implicit equation for the saddle point solution to solve for and then use a scaling form around the average value \eref{n_mean}, with the universal scaling $n=\log(t)+\gamma+\sigma_t\theta$, where $\theta$ is the unique parameter describing the distribution class in the large time limit. 
Instead of summing first over the $k_i$s in \eref{rho_n_int}, we can first perform the 
integrations on the $u_i$s, leading to the integral representation
\bb\label{rho_sum}
\rho_n(t)
=\frac{1}{n!}
\int_{-\pi}^{\pi}\frac{\drm 
y}{2\pi}
\e^{iy(t-n)}\left (
\sum_{l=0}^{t-n}\frac{\e^{-iyl}}{l+1} \right )^n
\ee
%
%
%
\begin{figure}
\centering
\includegraphics[scale=0.85,clip]{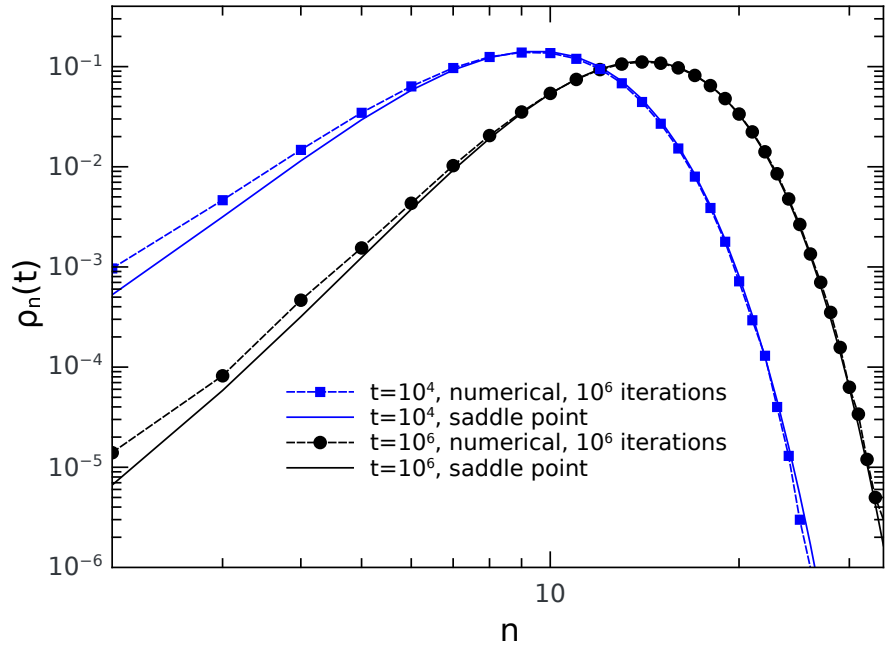}
\caption{\label{fig3a}
Distribution $\rho_n(t)$ of the number of fragments \eref{eq_as} at two given times $t=10^4$ and $t=10^6$, and comparison
with numerical results. The distribution shifts to higher values of $n$ as $t$ increases, since $\langle n\rangle _t\simeq \log(t)$, and is not symmetric, with higher probability for low values of $n$ than for higher $n$ above the average.}
\end{figure}
\begin{figure}
\centering
\includegraphics[scale=0.85,clip]{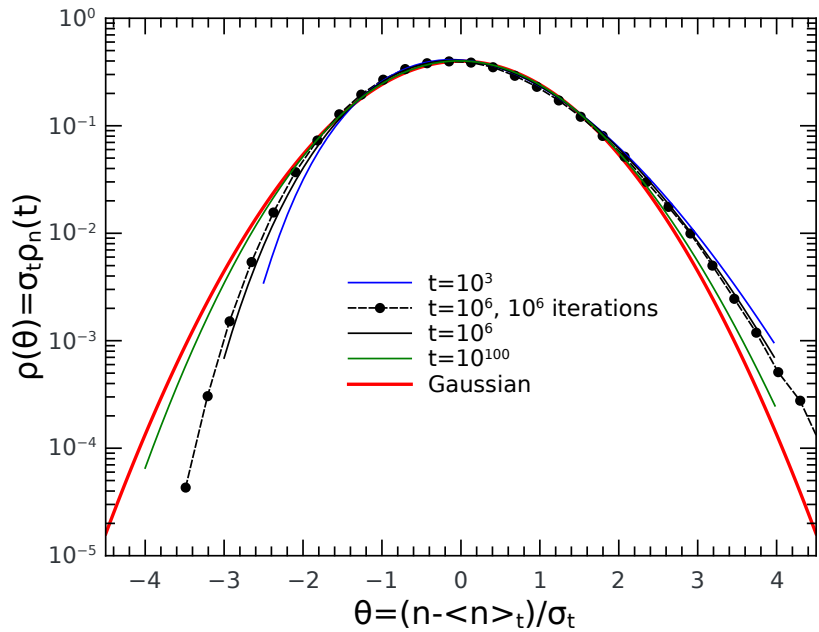}
\caption{\label{fig3b}Scaling distribution for different times. A comparison between numerical
simulation at $t=10^6$ and the saddle point solution for $\rho_n(t)$ \eref{eq_as} is displayed.
At extremely large times $t\gg 10^{100}$, the rescaled distribution approaches the Gaussian limit (red curve).}
\end{figure}
%
From the representation of the Kronecker delta function, we note that the upper limit
$t-n$ in the sum \eref{rho_sum} can be replaced by any larger finite value, since this does not modify the result and $t\gg n\sim \log(t)$ in the scaling limit. In the following we will take $t$ as the upper limit, avoiding the logarithmic singularity at $y=0$. We then apply the saddle point method to the argument function \cite{Temme:93}
\bb
\phi(y)=iy(t-n)+n\log\left ( \sum_{l=0}^{t}\frac{\e^{-iyl}}{l+1}\right )
\ee
The saddle point equation $\phi'(y)=0$ satisfies the equation
\bb\label{eq_saddle}
t=n\frac{\sum_{l=0}^{t}\e^{-iyl}}{\sum_{l=0}^{t}(l+1)^{-1}\e^{-iyl}},
\ee
When $t$ is large, the main contribution to 
the integral in \eref{rho_sum} comes from small $y$. We define the following scaling form for the saddle point solution: $y=i\xn(t)/t$, where $\xn(t)$ depends on $t$, with the asymptotic condition 
$\xn(t)/t\rightarrow 0$. Inserting this scaling relation in  \eref{eq_saddle}, 
and using results in  \ref{app0}, one obtains an implicit 
equation for $\xn$ for $n\ge 2$
\bb\label{eq_xn}
n\left . \frac{\e^{\xn}-1}{\xn} \right .=\log\left (t\right )+\gamma+\int_0^1\drm u\frac{\e^{\xn u}-1}{u},
\ee
There are two solutions, one unstable with $\xn\sim -t$, and one stable saddle point with $\xn>0$, when $n\ll t$. In particular the finite integral in \eref{eq_xn} has the asymptotic behaviors
\bb\nn
\int_0^1\drm u\frac{\e^{\xn u}-1}{u}\simeq -\log(-\xn)-\gamma,\;\xn\ll -1,
\\ 
\int_0^1\drm u\frac{\e^{\xn u}-1}{u}\simeq 
\exp(\xn)\left (\frac{1}{\xn}+\frac{1}{\xn^2}+\frac{2}{\xn^3}
+\cdots \right ),\;\xn\gg 1
\ee
Using results from \ref{app0}, one finds that the long-time asymptotic value of $\rho_n(t)$ is given when $n\ge 2$
by
\bb\label{eq_as}
\rho_n(t)\simeq\frac{1}{n!}\frac{\exp[\phi(i\xn /t)]}{\sqrt{-2\pi 
\phi''(i\xn/t)}}
\simeq
\frac{1}{n!}\frac{\e^{-\xn}\left [n(\e^{\xn}-1)/\xn \right ]^n}
{t\sqrt{2\pi[(1-\e^{-\xn})^{-1}-1/\xn-1/n]}}
\ee
with $\rho_1(t)=1/t$. In \efig{fig3a} we have compared the asymptotic limit \eref{eq_as} with the explicit solution as well as
the numerical simulations.
The scaling limit of the distribution $\rho_n(t)$ is performed by considering the typical value
of $n=\log(\tau)+\sigma_t\theta$ where $\tau=t\e^{\gamma}$ and $\theta$ the real parameter describing the limit
distribution around the average value. In this case, in \eref{eq_xn} the factor $n$ on the left hand side of the equation is comparable to $\log\tau$ on the right hand side, and the solution can be found by expanding the expression for small value of the continuous variable $\xn\rightarrow\alpha(\theta)$ which leads, up to the second order in $1/\sqrt{\log(\tau)}$, to
\bb\label{alpha}
\alpha(\theta)= -2\sigma_t\theta/\log\tau+\frac{(\sigma_t\theta)^2}{2\log(\tau)^2}+\cdots
\ee
Inserting only the dominant term of this expansion in \eref{eq_as}, as well as using the Stirling formula $\log n!\simeq n\log n-n+\log\sqrt{2\pi n}$, one finds, after some algebra, that the scaling limit of the distribution is given by a pure Gaussian function
\bb
\rho(\theta)=\lim_{t\rightarrow \infty}\sigma_t(t)\rho_n(t),\;
\simeq\frac{\e^{\gamma}}{\sqrt{\pi}}\frac{\exp\left (-\theta^2/2\right )}{\sqrt{2\pi}}
\ee
after keeping the terms proportional to $\sigma_t^2/\log(\tau)\simeq 1$ and discarding corrective terms in
$1/\sqrt{\log(\tau)}$.
The universal distribution is therefore given by a Gaussian process. The distribution is not exactly normalized, as there is an extra factor $\e^{\gamma}/\sqrt{\pi}\simeq 1.005$ which is however still close to unity, probably due to the fact we kept only the linear term in the solution \eref{alpha}. The logarithmic corrective terms are responsible for the non symmetric shape of the distribution, as seen in \efig{fig3b}. However, on this figure, the Gaussian limit is reached only for considerable long times, $t> 10^{100}$ unit steps, which are not accessible in numerical simulations, so that in appearance the distribution remains asymmetric and still far from the Gaussian limit.

\section{Fragment size distribution\label{sec_frag_size}}
%
In this section, we evaluate the probability distribution of the 
fragment sizes $x_i$ at a given time $t$. A power law distribution would indicate
a critical system for which all scales in the fragmentation process are present. 
To compute the fragment sizes, we first define the partial distribution $P_n(x,t)$ for $n+1$ fragments, using \eref{Q_gen}
\bb\fl\label{Puin}
P_{n}(x,t)=
\sum_{k_1+\cdots+k_n=t-n}
\int_0^1\drm u_1u_1^{k_1}\cdots
\int_0^{u_{n-1}}\drm u_{n}u_{n}^{k_{n}}
\left [
\frac{1}{n+1}
\sum_{i=0}^{n}\delta(x-u_i+u_{i+1})
\right ]
\ee
with $u_0=1$ and $u_{n+1}=0$. After integrating over the delta function, one has
\bb\fl\fl\nn
P_{n}(x,t)=\frac{1}{n+1}\sum_{i=0}^n\sum_{k_1+\cdots+k_n=t-n}
\int_x^1\drm u_1 u_1^{k_1}
\cdots \int_x^{u_{i-1}}\drm u_{i}u_{i}^{k_{i}}
(u_{i}-x)^{k_{i+1}}
\\ \times\label{eq_size_n}
\int_0^{u_i-x}\drm u_{i+2}u_{i+2}^{k_{i+2}}\cdots\int_0^{u_{n-1}}\drm u_{n}u_n^{k_n}
\ee
The total distribution $P_x(x,t)$ is then given by summing up over $n$, 
\bb
\Px(x,t)=\sum_{n=1}^tP_n(x,t)
\ee
The details of the calculations are given in \ref{app1}, where in particular 
the multiple integral is performed recursively. One finds that $P_n(x,t)$ is expressed 
as a partial derivative over $x$, see \eref{Pn}. Indeed, it can be written as
\bb\fl\fl\label{Px}
\Px(x,t)=\frac{1}{x}\frac{\partial}{\partial x}
\left [
-(1-x)^t\sum_{n=1}^t\frac{\rho_n(t)}{n+1}-
\sum_{r=1}^t\bin{t}{r-1}x^{t+1-r}(1-x)^r\sum_{n=1}^r\frac{\rho_n(r)}{n+1}
\right ]
\ee
Using the long time approximation \eref{Wilf_eq}, one finds that
\bb
\sum_{n=1}^t\frac{\rho_n(t)}{n+1}\simeq\int_0^1\drm u\frac{t^{u-1}}{\Gamma(u)}
\simeq\frac{1}{\log(t)}
\ee
Near $x=0$, one can approximate \eref{Px} by keeping the dominant terms in $x$ inside
the bracket, which corresponds to $r=t$, and therefore
\bb
\Px(x,t)\simeq \frac{t}{\log(t)}(1-x)^{t-1}\simeq \frac{t}{\log(t)}\e^{-xt}
\ee
This approximation is valid when $x<1/t$, beyond which one needs to study the 
other terms of the expression \eref{Px}. Using as before \eref{Wilf_eq}, one 
has
\bb\fl\nn
\Px(x,t)\simeq \int_0^1\frac{\drm u}{\Gamma(u)}
\frac{1}{x}\frac{\partial}{\partial x}
\left [-(1-x)^tt^{u-1}-\sum_{r=1}^t\bin{t}{r-1}x^{t+1-r}(1-x)^rr^{u-1}
\right ]
\\
\simeq \int_0^1\frac{\drm u}{\Gamma(u)}
\frac{1}{x}\frac{\partial}{\partial x}
\left [-(1-x)\sum_{r=0}^t\bin{t}{r}x^{t-r}(1-x)^r(r+1)^{u-1}
\right ]
\ee
where the terms proportional to $(1-x)^t$ was discarded for finite $x>1/t$. We 
can then express $(r+1)^{u-1}$ using an integral representation
\bb
\frac{1}{(r+1)^{1-u}}=\frac{1}{\Gamma(1-u)}\int_0^{\infty}\frac{\drm 
w}{w}w^{1-u}\e^{-(1+r)w} 
\ee
in order to perform the finite sum over $r$
\bb\fl
\Px(x,t)\simeq
 \int_0^1\frac{\drm u}{\Gamma(u)\Gamma(1-u)}
\frac{1}{x}\frac{\partial}{\partial x}
\left [-(1-x)\int_0^{\infty}\frac{\drm w}{w}
w^{1-u}\left \{
x+(1-x)\e^{-w}\right \}^t\e^{-w}
\right ]
\ee
In this expression, the dominant part of the integral over $w$ is near the origin
for large times, $w\simeq 0$, where an expansion gives directly
\bb\fl\label{eq_Px}
\Px(x,t)\simeq\frac{1}{tx(1-x)}
 \int_0^1\frac{\drm u}{\Gamma(u)}
ut^u(1-x)^u\simeq \frac{1}{x\log[t(1-x)]},\;t\gg 1
\ee
%
\begin{figure}
\centering
\includegraphics[scale=0.85,clip]{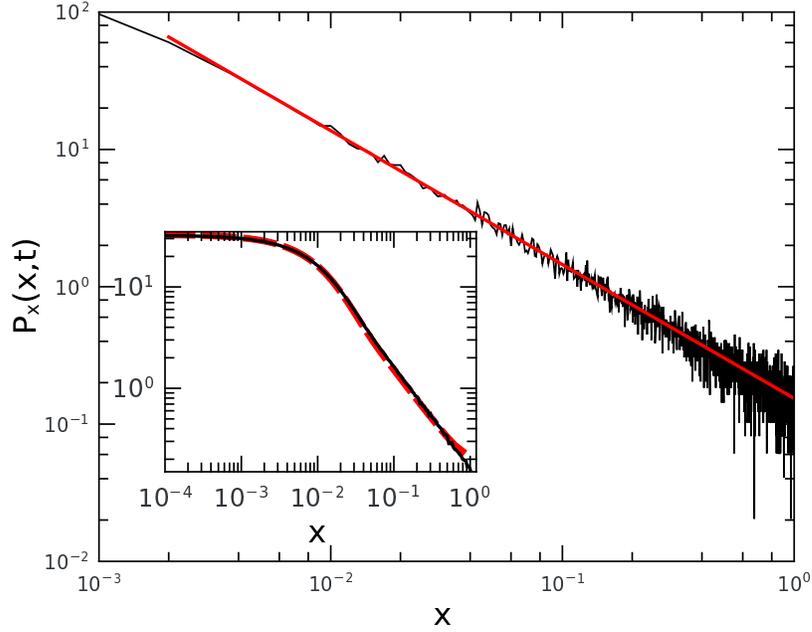}
\caption{\label{fig4} Level size distribution for $t=10^7$ and $10^5$ samples (black line). 
The fit (red line) gives a power law with an exponent equal to $-0.975\pm 0.002$ close to $-1$.
Insert: distribution calculated from \eref{Px} (red dashed line), 
and result from numerical simulations (black line) at $t=100$ for $10^7$ samples.}
\end{figure}
%
We could expand the function $1/\Gamma(u)$ using the $\gamma_i$ coefficients and integrate over $u$, but
the dominant contribution of the integral is for $u$ close to unity. Therefore the distribution decays
like $1/x$ for most part of the interval $x>1/t$, with corrective terms in powers of $1/\log[t(1-x)]$.
Figure \ref{fig4} gives the comparison between numerical data and formula \eref{Px}, for which the 
local distributions $\rho_n(r)$ are approximated by \eref{Wilf_eq}. In the sum over $r$ in \eref{Px}, this approximation is not accurate for integers $r$ small, but we assume that these terms does not contribute to the sum. 
We can compare this distribution to a simple problem of partition of the unit segment with a set of $n$ random numbers
$0<x_1<x_2<\cdots <x_n<1$, taken each from a uniform distribution and ordered by increasing values. We also take variable $n$ for each set from an exponential distribution $(1-\e^{-a})\e^{-a(n-1)}$, with $a>0$. After some  algebra, one finds that the size distribution for each set $n$ is given by $P_n(x)=n(1-x)^{n-1}$, and that the global
size distribution $\Px(x)$, after summation over $n$, is simply
\bb
\Px(x)=\frac{1-\e^{-a}}{\left [1-(1-x)\e^{-a} \right ]^2}
\ee
To select partitions with a maximum of fragment numbers, we consider the limit where the parameter $a$ is small, for which we can approximate $\Px(x)\simeq ax^{-2}\e^{-2a(1-x)/x}$ when $x>a$. The size distribution for this simple model is then given by a power law with exponent -2 instead of -1 as in \eref{eq_Px}. Similarly parameter $a$ can be thought as a time variable $a\simeq \log(t)$, with the probability of having a set of $n+1$ fragments increasing with time. In both cases a universal power law distribution is commonly found with corrective terms depending on the details of the process.
%
%
\section{Stochastic process on a semi-infinite chain}
%
We consider in this section the fragmentation/recombination process on a 
semi-infinite chain$[0,\infty]$. The probability of cutting the chain
at position $x$ is given by an exponential decay $p(x)=\lambda\e^{-\lambda 
x}$, where $\lambda$ is a positive constant. One starts with the semi-infinite
axis as initial condition, and implements the stochastic 
process displayed on \efig{fig_process}. The time-dependent master equation
can be written similarly as \eref{eq_master}
\bb\nn\fl
P^{(n)}_{t+1}(\EE_0,\cdots,\EE_n)=
\lambda\exp[-\lambda(\EE_0+\cdots+\EE_n)]\Big \{
P^{(n-1)}_{t}(\EE_0,\cdots,\EE_{n-2},\EE_{n-1})
\\ \fl\fl\label{eq_master_inf}
+\int_{\EE_{n}}^{\infty}\drm\EE'_{n}P^{(n)}_{t}(\EE_0,\cdots, 
\EE_{n-1},\EE'_{n})
+\int_{\EE_{n}}^{\infty}\drm\EE'_{n}
\int_{0}^{\infty}\drm\EE'_{n+1}
P^{(n+1)}_{t}(\EE_0,\cdots,\EE_{n-1},\EE'_{n},\EE'_{n+1})+\cdots \Big \}
\ee
Initial condition is given by the
distribution $P_0^{(0)}(\EE_0)=\lambda\e^{-\lambda \EE_0}$, after starting 
with a semi-infinite interval and choosing a random point 
$\EE_0$ to construct the first fragment $[0,\EE_0]$ with the previous 
probability. This will define the ground state of the system. Equation 
(\ref{eq_master_inf}) can be solved recursively and uniquely, by considering  
the set of new variables $\EF_i=\e^{-\lambda \EE_i}$, and the scaling form
\bb
P^{(n)}_{t}(\EE_0,\cdots,\EE_n)=(\lambda^{n+1}\EF_0\cdots\EF_n)Q^{(n)}_t(\EF_0,
\cdots,\EF_n)
\ee
This allows to simplify the master equation \eref{eq_master_inf}
\bb\fl\nn
Q^{(n)}_{t+1}(\EF_0,\cdots,\EF_n)=
(\EF_0\cdots\EF_{n-1})\Big \{
Q^{(n-1)}_{t}(\EF_0,\cdots,\EF_{n-2},\EF_{n-1})
\\ \fl\fl\label{eq_master_inf_Q}
+\int_{0}^{\EF_n}\drm\EF'_{n}Q^{(n)}_{t}(\EF_0,\cdots, 
\EF_{n-1},\EF'_{n})
+\int_{0}^{\EF_n}\drm\EF'_{n}
\int_{0}^{1}\drm\EF'_{n+1}
Q^{(n+1)}_{t}(\EF_0,\cdots,\EF_{n-1},\EF'_{n},\EF'_{n+1})
+\cdots \Big \}
\ee
The general solution of this time-dependent equation with initial condition 
$Q_0^{(0)}=1$ is given in terms of polynomial products only
\bb
Q^{(n)}_{t}(\EF_0,\cdots,\EF_n)=(\EF_0^t\EF_1^{n-1}\EF_2^{n-2}\cdots\EF_{n-1}
^1)\times
\sum_{k_1=0}^{t-n}\EF_1^{k_1}\sum_{k_2=0}^{k_1}\EF_2^{k_2}\cdots\sum_{k_n=0}^{k_
{n-1}}\EF_{n}^{k_n}
\ee
Hence
\bb\label{sol_inf}
P^{(n)}_{t}(\EE_0,\cdots,\EE_n)=(\lambda^{n+1}\EF_0^{t+1}\EF_1^{n}\EF_2^{n-1}
\cdots\EF_{ n-1 }^2\EF_n^1)\times
\sum_{k_1=0}^{t-n}\EF_1^{k_1}\sum_{k_2=0}^{k_1}\EF_2^{k_2}\cdots\sum_{k_n=0}^{k_
{n-1}}\EF_{n}^{k_n}
\ee
The distribution of fragments $\de_n(t)$ as function of time is given by the 
integrals
\bb\nn
\de_n(t)=\int_0^{\infty}\drm\EE_0\int_0^{\infty}\drm\EE_1\cdots\int_0^{\infty}
\drm\EE_nP^{(n)}_{t}(\EE_0,\cdots,\EE_n)
\\ \nn
=\int_0^{1}\drm \EF_0\int_0^{1}\drm \EF_1\cdots\int_0^{1}
\drm \EF_nQ^{(n)}_{t}(\EF_0,\cdots,\EF_n)
\\ \label{rho_def2}
=\frac{1}{1+t}\sum_{k_1=0}^{t-n}\frac{1}{n+k_1}\sum_{k_2=0}^{k_1}\frac{1}{
n-1+k_2 } 
\cdots\sum_{k_{n-1}=0}^{k_{n-2}}\frac{1}{2+k_{n-1}}\sum_{k_n=0}^{k_{n-1}}\frac{1
} { 1+k_n } 
\ee
Comparing this expression with \eref{Q_gen} and \eref{rho_def}, these expressions
are equivalent. Indeed, we can reformulate \eref{rho_def} by integrating 
directly on variables $u_i$ without using the symmetrization with respect to 
the $u_i$s in the finite interval case, which is not necessary in \eref{sol_inf}, such that
\bb\nn
\rho_n(t)=
\sum_{k_1=0}^{t-n}
\int_0^1\drm u_1 u_1^{t-n-k_1}
\sum_{k_2=0}^{k_1}
\int_0^{u_1}\drm u_2u_2^{k_1-k_2}\cdots
\\ \nn
\sum_{k_{n-2}=0}^{k_{n-3}}
\int_0^{u_{n-3}}\drm u_{n-2}
u_{n-2}^{k_{n-3}-k_{n-2}}
\sum_{k_{n-1}=0}^{k_{n-2}}
\int_0^{u_{n-2}}
u_{n-1}^{k_{n-2}-k_{n-1}}
\int_0^{u_{n-1}}\drm u_nu_n^{k_{n-1}}
\\ 
=\frac{1}{t}\sum_{k_1=0}^{t-n}\frac{1}{n-1+k_1}\sum_{k_2=0}^{k_1}\frac{1}{
n-2+k_2 } 
\cdots\sum_{k_{n-1}=0}^{k_{n-2}}\frac{1}{1+k_{n-1}}=\de_{n-1}(t-1)
\ee
The two distributions are equal, up to a shift in indices due to a different 
convention for the origin of time and number of fragments (we could have used 
instead the initial condition $P_0^{(0)}=1$).
%
%
\section{Two-time auto-correlation function\label{sec_correlation}}
%
In this section, we analyse the two-time auto-correlation functions using exact solutions of the dynamics. We define here the auto-correlation functions as the correlation between two configurations with the same fragment number at different times $t>s$. To be more precise, we first consider a given configuration of $(m+1)$ intervals $(v_1,\cdots,v_m)$ at time $s$, with $v_1\ge v_2\ge v_3\ge \cdots \ge 
v_m>0$ and probability $Q_s^{(m)}(v_1,\cdots,v_m)$. 
A quantum representation of this state is given by the vector
$|v_1,v_2,\cdots,v_m;s\rangle$. Starting with the initial vector $|0\rangle$ representing the initial
fragment of length unity, one can rewrite the probability $Q_s^{(m)}(v_1,\cdots,v_m)$
as the amplitude of transition between states $|0\rangle$ and 
$|v_1,v_2,\cdots,v_m;s\rangle$  
\bb
Q_s^{(m)}(v_1,\cdots,v_m)=\langle v_1,\cdots,v_m;s|0\rangle
\ee
The auto-correlation function $C_d(t,s)$ (disconnected part) is defined here as the probability of having the same number of fragments at times $t$ and $s$. This is represented by the summation (there are at most $s+1$ fragments at time $s$)
\bb\nn
C_d(t,s)=\int\drm u\int\drm v\sum_{m=1}^s
\langle u_1,\cdots,u_m;t|v_1,\cdots,v_m;s\rangle\langle v_1,\cdots,v_m;s|0\rangle
\\ \label{Ctsd_def}
=\int\drm u\int\drm v\sum_{m=1}^s
\langle u_1,\cdots,u_m;t|v_1,\cdots,v_m;s\rangle Q_s^{(m)}(v_1,\cdots,v_m)
\ee
We then defined the connected part as
\bb \label{Cts_def}
C(t,s)=C_d(t,s)-\sum_{m=1}^{s}\rho_m(s)\rho_m(t)
\ee
In order to find the intermediate amplitude $\langle u_1,\cdots,u_m;t|v_1,\cdots,v_m;s\rangle $ one has to apply the equation of evolution \eref{eq_master2} at successive times, from $s$ to $t$, starting with the condition
\bb\label{init_condQ}
\langle u_1,\cdots,u_m;s|v_1,\cdots,v_m;s\rangle=\prod_{i=1}^m\delta(u_i-v_i)=\Delta_m
\ee
The evolution of this initial condition is constrained by the master 
equation \eref{eq_master}. We will note for simplification
\bb\label{def_Qts}
Q_{t,s}^{(n)}(u_1,\cdots,u_n)=\langle u_1,\cdots,u_n;t|v_1,\cdots,v_m;s\rangle
\ee
Using the results of \ref{app2}, at time $t=s+l$ and for $m\ge l$, one finds that 
\begin{equation}\nn
\left \{
\begin{array}{l}
Q_{s+l,s}^{(m+l)}(u_1,\cdots,u_{m+l})=\Delta_m,
\\ \nn
Q_{s+l,s}^{(m+l-1)}(u_1,\cdots,u_{m+l-1})=\Delta_{m-1}
+\Delta_m(u_{ m+1}+\cdots+u_{m+l-1}),
\\ \nn
Q_{s+l,s}^{(m+l-2)}(u_1,\cdots,u_{m+l-2})=\Delta_{m-2}
+\Delta_{m-1}(u_{ m}+\cdots+u_{m+l-2})
\\ \nn
+\Delta_{m}\sum_{k_1+\cdots +k_{l-2}=2}u_{m+1}^{k_1}\cdots 
u_{m+l-2}^{k_{l-2}},
\\ \nn
\cdots 
\\ \nn
Q_{s+l,s}^{(m)}(u_1,\cdots,u_{m})=\Delta_{m-l}+\Delta_{m-l+1}(u_{m-l+2}+\cdots+
u_{m})+\cdots
\\ \nn
+\Delta_{m-2}\sum_{k_0+k_1=l-2}u_{m-1}^{k_1}u_{m}^{k_0}
+\Delta_{m-1}u_m^{l-1},
\\ \nn
\cdots 
\\ \label{Qcorr}
Q_{s+l,s}^{(1)}(u_1)=\Delta_0 u_1^{l-1}
\end{array}
\right .
\end{equation}
The solutions for the first increments in time are given in \ref{app2}. Symbols $\Delta_k$ are 
short notation for $\Delta_{m-1}= \prod_{i=1}^{m-1}\delta(u_i-v_i)\theta(u_{m}-v_{m})$, 
$\Delta_{m-2}=\prod_{i=1}^{m-2}\delta(u_i-v_i)\theta(u_{m-1}-v_{m-1})$, and so on. The $\theta$ functions
are important here to select the correct ordering between $u_i$ and $v_i$.
We can check that these probabilities are normalized at each time step
\bb
\sum_{r=1}^{m+l}\int_0^1\drm u_1\int_0^{u_1}\drm u_2\cdots \int_0^{u_{r-1}}\drm 
u_r
Q_{s+l,s}^{(r)}(u_1,\cdots,u_r)=1
\ee
The auto-correlation \eref{Cts_def} is similar to the usual two-time correlation function 
$C({\bf r'},t;{\bf r},s)$ defined in physical systems on a lattice and taken at the same point in space
${\bf r}={\bf r'}$ (and averaged over ${\bf r}$). The distance $|{\bf r-r'}|$ between two points may be identified as the difference $|m-n|$ between two configurations $(u_1,\cdots,u_n)$ and $(v_1,\cdots,v_m)$. And the sum over all possible $m$ in \eref{Ctsd_def} is equivalent to averaging over all the positions in space $C(t,s)=\sum_{\bf r}C({\bf r},t;{\bf r},s)$. The disconnected part of the auto-correlation function \eref{Ctsd_def} can be computed exactly
and expressed into discrete sums. In particular, $C_d(t,s)$ can be expressed as an expansion in $(s/t)$, and the first terms are given by products of the density function (see \ref{app2} and \ref{app3} for technical details)
\bb\fl\fl\label{Ctsd_sol}
C_d(t,s)=\sum_{m=1}^s\left \{
\rho_m(s)\rho_m(t-s)+\frac{s-1}{t-1}\rho_m(s-1)[\rho_{m-1}(t-s)-\rho_m(t-s)]+\cdots
\right \}
\ee
and therefore
\bb\fl\fl\label{Cts_sol}
C(t,s)=\sum_{m=1}^s\left \{
\rho_m(s)[\rho_m(t-s)-\rho_m(t)]+\frac{s-1}{t-1}\rho_m(s-1)[\rho_{m-1}(t-s)-\rho_m(t-s)]+\cdots
\right \}
\ee
%
\begin{figure}
\centering
\includegraphics[scale=0.9,clip]{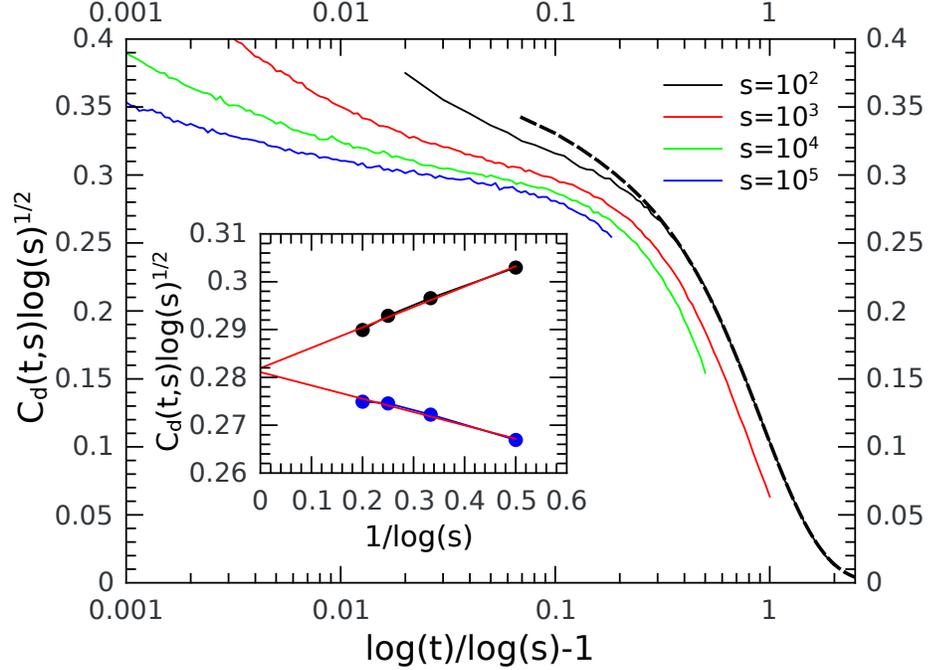}
\caption{\label{fig_corr0}Two-time probability function $C_d(t,s)$ for several values of $s=10^2,10^3,10^4$ and $10^5$. The number of samples used for each value of $s$ is $10^6$. The dashed
line is the approximation \eref{Ctsd_sol} for $s=10^2$. Inset: plot of $C_d(t,s)\sqrt{\log(s)}$ as function of $1/\log(s)$ for $t/s=2$ (black dots) and $t/s=4$ (blue dots). The linear fits (red lines) are in agreement with the limit $C_d(t,s)\sqrt{\log(s)}\simeq (4\pi)^{-1/2}\simeq 0.2821$ when $s\rightarrow\infty$ at fixed $t/s$.}
\end{figure}
%
The first dominant term in \eref{Ctsd_sol} is the products of the probabilities to have $(m+1)$ fragment both at times $s$ and $t-s$, while the other contributions appear as higher powers of $(s/t)$ which are negligible when $t\gg s\gg 1$, see \ref{app3}. Using \eref{Wilf_eq}, we can compute numerically the dominant term of \eref{Ctsd_sol} and its first correction in $(s/t)$, and compare them with the numerical data for which $10^6$ samples are used,
see \efig{fig_corr0}. In this figure, four different values of $s=10^2,10^3,10^4$ and $10^5$ are used. There is a very good agreement with \eref{Ctsd_sol} for $t/s>2$. The corrections to the dominant contribution appear when $t/s<2$, and \eref{m_sup_l} has to be taken into account (see \ref{app3} for details), for which fragments from the initial condition are not completely erased or replaced.
Otherwise the aging phenomena can be seen clearly as the different curves show a strong dependence on the initial time $s$. The dominant part of $C_d(t,s)$ can be written as
\bb\label{Cts_app}
C_d(t,s)\simeq \frac{1}{s(t-s)}\sum_{m=1}^s\sum_{i,j=1}^m\gamma_i\gamma_j
\frac{[\log(s)]^{m-i}}{(m-i)!}
\frac{[\log(t-s)]^{m-j}}{(m-j)!}
\ee
This formula is accurate to describe the long time behavior of the auto-correlation function. In the large
$s$ limit, we assume that the finite sum over $m$ can be transformed into a series with the upper limit $s$ replaced by $\infty$, since the different terms of the sum should decrease fast enough.
In particular, one can express all the terms with modified Bessel functions, using $\xi=\log(s)\log(t-s)$ 
\bb\nn
C_d(t,s)\simeq \frac{1}{s(t-s)}\left [
\sum_{i\ge 1}\gamma_i^2I_0(2\sqrt{\xi})
\right .
\\ 
+\label{Cts_series}
\left .
\sum_{1\le i<j}\gamma_i\gamma_j
\left (\frac{1}{\log(s)^{j-i}}+\frac{1}{\log(t-s)^{j-i}}\right )
\xi^{j-i}\frac{\partial^{j-i}}{\partial\xi^{j-i}}I_0(2\sqrt{\xi})
\right ]
\ee
Keeping the dominant terms of the Bessel derivatives, one finds the following
expression
\bb\fl\fl\nn
C_d(t,s)\simeq\frac{1}{s(t-s)}\left [
\gamma_1^2I_0(2\sqrt{\xi})+
\gamma_1\gamma_2\left (
\sqrt{\frac{\log(t-s)}{\log(s)}}+\sqrt{\frac{\log(s)}{\log(t-s)}}
\right )I_1(2\sqrt{\xi})
\right .
\\ \label{Ctsd_Bessel0}
\left .
+\gamma_1\gamma_3\left (
\frac{\log(t-s)}{\log(s)}+\frac{\log(s)}{\log(t-s)}
\right )I_2(2\sqrt{\xi})
+\gamma_2^2I_0(2\sqrt{\xi})+\cdots
\right ]
\ee
The index of the Bessel functions corresponds to the difference of index between the $\gamma_i$ coefficients.
Let us consider the scaling limit $\log(s)\gg 1$ with $t/s$ fixed. One can expand $C_d(t,s)$ using
the asymptotic limit of the Bessel functions $I_n(2\sqrt{\xi})\simeq  \e^{2\sqrt{\xi}}(4\pi)^{-1/2}\xi^{-1/4}$. Also, the ratios $\log(s)/\log(t-s)$ or $\log(t-s)/\log(s)$ are close to unity, up to corrections in $1/\log(s)$, which simplifies further the sum, and therefore
\bb\fl\label{Ctsd_lim}
C_d(t,s)\simeq \frac{1}{s(t-s)}\left [
\sum_{i<j}2\gamma_i\gamma_jI_{j-i}(2\sqrt{\xi})+\sum_i\gamma_i^2I_0(2\sqrt{\xi})
\right ]
\simeq \frac{1}{s(t-s)}\frac{\e^{2\sqrt{\xi}}}{\sqrt{4\pi}\xi^{1/4}}
\ee
after recognizing the perfect square $(\sum_i\gamma_i)^2=1$. Setting $u=t/s$, considered here as a fixed ratio,
we can expand for large $s$ the argument $\sqrt{\xi}\simeq \log(s)+\ff\log(u-1)$, and after further simplifications one finds the following limit
\bb\label{Ctsd_scaling}
C_d(t,s)\sqrt{\log(s)}\simeq (4\pi)^{-1/2}\simeq 0.2821
\ee
which is independent of $t/s$. This scaling limit was checked numerically in the inset of \efig{fig_corr0}, for
$t/s=2$ and $t/s=4$. The extrapolation of the fits towards the origin $1/\log(s)\rightarrow 0$ gives the correct numerical constant. This explains the choice of the scaling in \efig{fig_corr0} for which the auto-correlation function is approximately proportional to $\log(s)^{-1/2}$. The connected part of the correlation function $C(t,s)$ \eref{Cts_sol} is approximately equal to $C(t,s)\simeq\sum_{m=1}^s\rho_m(s)[\rho_m(t-s)-\rho_m(t)]$, which can be evaluated by subtracting the expression \eref{Ctsd_Bessel0} or \eref{Ctsd_lim} function of $\xi$ by the same expression function of $\xi=\log(s)\log(t)$ instead
\bb\nn\fl
C(t,s)&\simeq& \frac{1}{\sqrt{4\pi}s^2}\left (
\frac{1}{u-1}\frac{\exp[2\log(s)\sqrt{1+\log(u-1)/\log(s)}]}{\log(s)^{1/2}[1+\log(u-1)/\log(s)]^{1/4}}
\right .
\\ \nn
& & \left .
-\frac{1}{u}\frac{\exp[2\log(s)\sqrt{1+\log(u)/\log(s)}]}{\log(s)^{1/2}[1+\log(u)/\log(s)]^{1/4}}
\right )
\\ \label{Cts_lim}
& &\simeq\frac{1}{8\sqrt{\pi}\log(s)^{3/2}}\left [\log(u)(\log(u)+1)-\log(u-1)(\log(u-1)+1)\right ]
\ee
The autocorrelation function is positive when $u>2\e^{-1}/(\sqrt{1+4/e}-1)\simeq 1.286$ and negative otherwise. The 
other corrective terms are smaller than $1/\log(s)^{3/2}$ and are not taken into account in this analysis.

\section{Modified fragmentation process with reset\label{sec_reset}}
%
In this section, we introduce a probability $r$ for the system to reset to its initial condition,
which is a single fragment of unit length. The idea is to see how the aging process seen in
the previous section is perturbed by a regular reset, with the introduction of a reset time scale $1/r$. This process was introduced for example in coagulation-diffusion processes to probe the non-trivial effects on correlation functions depending on the different length scales \cite{Durang:2014}. The introduction of the probability parameter $r$ modifies slightly \eref{eq_master2} by adding a contribution $Q_t^{(0)}$ from the presence of single segments and new terms coming from previous resets
in $Q_t^{(n)}$
\bb\nn
Q_t^{(0)}=r,
\\ \nn
Q_t^{(n)}(u_1,\cdots,u_n)=(1-r)^t
\sum_{\sum_i k_i=t-n}(u_1^{k_1}\cdots u_n^{k_n})
\\ \label{Q_gen_r}
+r(1-r)^{t-1}\sum_{\sum_{i}k_i=t-n-1}(u_1^{k_1}\cdots u_n^{k_n})
+\cdots+r(1-r)^n
\ee
As before, we can check that the total distribution at time $t$ is normalized
\bb
\sum_{n=0}^{t}\int_0^1\drm u_1\int_0^{u_1}\drm u_2\cdots \int_0^{u_{n-1}}
\drm u_n Q_t^{(n)}(u_1,\cdots,u_n)=1
\ee
The distribution of the fragment number $\rho_{n,r}(t)$ with reset $r$ can be evaluated
using a Kronecker integral, which allows for the summation over all the $k_i$s.
One obtains
\bb\fl\fl\nn
\rho_{n,r}(t)=\frac{1}{n!}\int_{-\pi}^{\pi}\frac{\drm y}{2\pi}
\left [-\log(1-\e^{-iy}) \right ]^n\e^{iyt}
\left \{
(1-r)^t+r(1-r)^{t-1}\e^{-iy}+\cdots+r(1-r)^n\e^{-iy(t-n)}
\right \}
\ee
The logarithmic function can be expanded using the series \eref{def_stirling}, and
the different integrations over $y$ inside the sum lead to the expression
\bb\label{rho_nr}
\rho_{n,r}(t)=
(1-r)^t\rho_n(t)+r\sum_{p=n}^{t-1}(1-r)^p\rho_n(p)
\ee
%
\begin{figure}
\centering
\includegraphics[scale=0.9,clip]{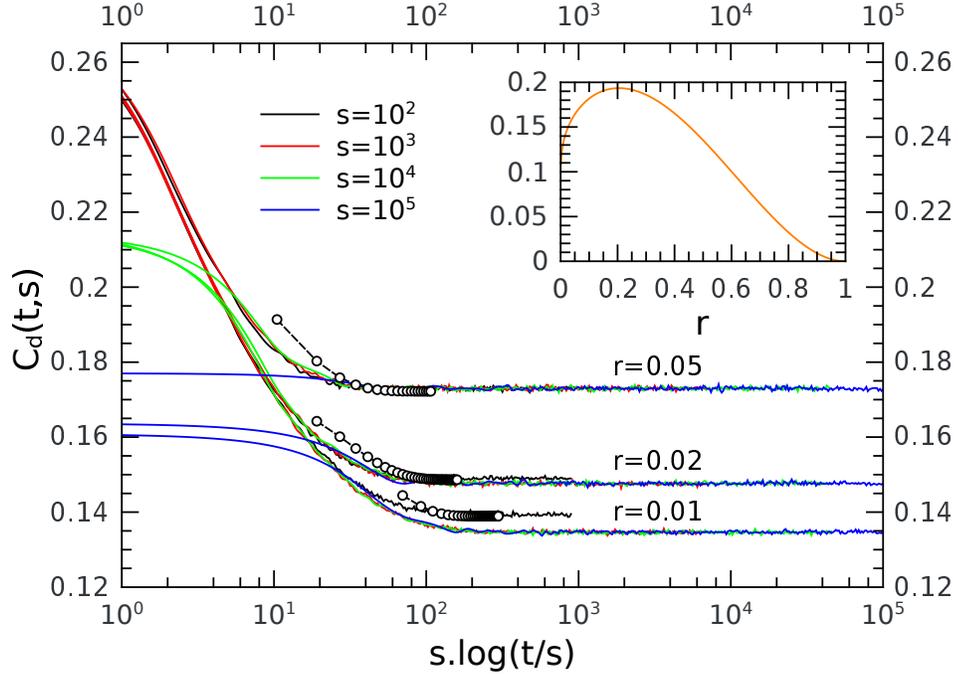}
\caption{\label{fig_corr_r}
Auto-correlation function $C_d(t,s)$ for several values of $r$ and $s$. Contrary to the case without reset
$C_d(t,s)$ does not depend on $s$ for $t>1/r$ sufficiently large. After an initial
decay, all auto-correlation functions tend to a constant depending on $r$ and given approximately by the Bessel function in \eref{Ctsd_r}. The symbol curves represent the formula $C_d(t,s)=\sum_{m=1}^{s}\rho_{m,r}(s)\rho_{m,r}(t-s)$ with $\rho_{m,r}(t)$ computed from \eref{rho_nr} and \eref{Wilf_eq} and $s=100$. This approximation fits well the data.
Inset: Approximate asymptotic value of $C_d(t,s)$ \eref{Ctsd_Bessel} as function of $r$. A maximum auto-correlation is reached for $r\simeq 0.2048$ for which $C_d(t,s)\simeq 0.1935$.}
\end{figure}
%
The asymptotic limit $t\rightarrow\infty$ with $r$ finite corresponds to a time-independent
distribution, which is simply a Poisson distribution of constant rate equal to $-\log(r)$
\bb
\rho_{n,r}(t)\simeq r\sum_{p=n}^{\infty}(1-r)^p\rho_n(p)
=r\frac{[-\log(r)]^n}{n!}
\ee
Using the results of \ref{app2}, the dominant term of the auto-correlation function $C_d(t,s)$, in
the limit of large times, $t\gg s\gg 1$, can be successively approximated by
\bb\label{Ctsd_r}
C_d(t,s)&\simeq& \sum_{m=1}^{s}\rho_{m,r}(s)\rho_{m,r}(t-s)
\\ \label{Ctsd_Bessel}
&\simeq& \sum_{m=1}^{\infty}\frac{r^2}{(m!)^2}\left (-\log r \right )^{2m}
=r^2\left [
I_0(-2\log r)-1 \right ]
\ee
which is time-independent. In \efig{fig_corr_r} we have plotted $C_d(t,s)$ for different values of $r$ and $s$ as
function of $s\log(t/s)$ which was chosen for convenience. One observes that $C_d(t,s)$ does 
not depend on $s$ as $r$ increases, which is 
characteristic of a non-aging process, and the presence of fast relaxation effects. The first approximation \eref{Ctsd_r} with \eref{rho_nr} gives the correct result in the large time regime, see the symbol lines in \efig{fig_corr_r} for comparison, and saturates at a finite value. For example, for $r=0.01$, $C_d(t,s)\simeq 0.1333$ if we consider the second expression with the Bessel function in \eref{Ctsd_Bessel}.
\section{Discussion}
\com{
In this manuscript we have evidenced aging phenomena in a modified
fragmentation model that incorporates a process that slows the typical dynamics of producing
a steady large number of fragments, in a similar way that energy states relax slowly to equilibrium in glassy systems from an initial quenched state. This model without Hamiltonian and temperature displays characteristics of a slowing down process. The "quakes" are defined in general as rare events for which the number of fragments drops suddenly by a large amount. This is similar, in a free energy landscape description, to the activation process that takes the system from one valley to the other through large barriers. Activation process here is played by the stochastic mechanism.
The limit distribution of the fragment number is only reached for very long times, out of range of numerical simulations, and is Gaussian. The fragment size distribution also displays in this limit scale invariance as it follows an inverse 
power law for sizes $x$ larger than the threshold scale $x>1/t$. To reach such state, made of fragments of infinitely small sizes near the origin, a record process is necessary, since the small fragments on the left hand side have more chance to survive than fragments on the right hand side. Cutting the interval at the leftmost location, hence producing a "quake", will make the first fragment even smaller, as seen in \efig{fig_frag}, which is necessary to reach at later times a higher density of fragments on the left side. A "quake" is equivalent to a reconfiguration process that takes the system from one "energy valley" to another with a lower energy. 
This mechanism of our model generates also a slow logarithmic increase of the average number of fragments. The probability to reach such configuration of 
only two fragments decreases as $1/t$, since the accumulation of small fragments near the origin increases with time. For example, the probability distribution of the first fragment size $x_0$ at time $t$ is equal to $t(1-x)^{t-1}$, with a size average equal to $1/(t+1)$. Therefore
a new two-fragment configuration, or "record", is more and more difficult to obtain because the size of the first fragment has to be smaller and smaller. This model is an example of such record dynamics. The other fragments above the first one $x_0$ can be seen as fluctuations in this valley because they have a finite survival time.
There is also the question of "equilibrium state" in the model. In a mathematical sense, one may argue that it is defined by an infinite number of infinitesimally small fragments located near the origin and the system is driven slowly towards this state. One may also try to define an effective energy for this model, as $\log(x_0)<0$ can be thought as the bottom energy of the valley, which decreases with time. And the other fragment contributions can be seen as energy fluctuations with a hierarchical set of relaxation times, the last fragment having the shortest survival time. 
 More importantly the correlation functions as defined in part \ref{sec_correlation} show clear aging properties due to the slow relaxation.
The scaling behavior of the correlation function is found by considering, as before, the limit $s\rightarrow\infty$ with $t/s$ fixed. In the scaling theory of aging \cite{book:henkel_aging}, it is usually written, for auto-correlation functions, as}
\bb
C(t,s)\simeq s^{-b}F(t/s)
\ee
with $b$ the scaling exponent and $F$ the scaling function. Here clearly, one 
can instead identify from
 \eref{Cts_lim} the scaling relation $C(t,s)\simeq \log(s)^{-b}F(t/s)$, with a 
logarithmic form and $b=3/2$. The scaling function
$F(u)$ goes to zero as expected when $u\rightarrow\infty$, in particular $F(u)\simeq [1+2\log(u)]/u$.

The response function $R(t,s)$ to an external perturbation 
follows in general a similar scaling relation $R(t,s)=s^{-1-a}G(t/s)$ than the auto-correlation function, with $a$ a second scaling exponent, but this would have to be defined properly in the context of this fragmentation model, in contrast to the auto-correlation function which is more straightforward.
For example, in a coagulation-diffusion process in one 
dimension \cite{Durang:2011}, the response function is defined by inserting a 
particle on a given site at time $s$ as an external perturbation, and computing 
the probability at time $t$ that a particle is present on the same site. Here we 
may think of inserting an additional segment at time $s$ (of random size and 
constrained to be inside the unit interval), and compute the probability that at 
time $t$ the same number of segments is found. \com{ This study would then be relevant 
in the context of aging models. Also it would be interested to study the time of survival for a given fragment before it is removed by fragmentation, or the characteristic times between "quakes". }

The same process with reset $r$ as studied in part \ref{sec_reset} shows a different class of relaxation process, with apparently a conventional exponential decreasing and no more aging. Indeed one can easily identify from \efig{fig_corr_r} the length scale $t\simeq 1/r$ beyond which the system loses its memory and becomes Markovian. The correlation functions do not depend on $s$ anymore and reach a constant in the saturated regime which is maximum for a finite value of $r$.

\ack
I would like to acknowledge S. Boettcher for his useful comments on this manuscript.

%
\appendix
%
\section{\label{app0}}
In this appendix, different sums are evaluated in the limit of large $t$.
The saddle point equation \eref{eq_saddle} can be written as $t=nI_1/I_0$ 
where $I_0=\sum_{l=0}^t(l+1)^{-1}\e^{-iyl}$ and $I_1=\sum_{l=0}^t\e^{-iyl}$. If $y=i\xn(t)/t$, with 
$\xn(t)/t\rightarrow 0$ when $t$ is asymptotically large, $I_0$ is logarithmically diverging, and we 
regularize the sum by substracting the diverging part
\bb\nn
I_0=\sum_{l=0}^t\frac{\e^{-iyl}}{l+1}=\sum_{l=0}^t\frac{\e^{\xn l/t}-1}{l+1}+\sum_{l=0}^t\frac{
1}{l+1}
\simeq \sum_{l=0}^t\frac{\e^{\xn l/t}-1}{l+1}+\log(t)+\gamma
\ee
In this expression, one can replace $l/t$ by a continuous variable $u$ between 
0 and 1, such that
\bb\nn
I_0\simeq \log(t)+\gamma
+\int_{0}^1\drm u\frac{\e^{\xn u}-1}{u}
\ee
%
The other sum $I_1$ is asymptotically equivalent to
\bb\nn
I_1\simeq t\int_0^1\drm u\e^{\xn u}\simeq t(\e^{\xn}-1)/\xn
\ee
Combining $I_0$ 
and $I_1$ in \eref{eq_saddle}, one obtains the implicit equation \eref{eq_xn}. The second 
derivative of $\phi$ is equal to 
\bb
\phi''(y)=-\frac{n}{I_0^2}\left ( I_2I_0+I_1I_0-I_1^2 \right )
\ee
where $I_2=\sum_{l=0}^tl\e^{-iyl}$. We can estimate asymptotically the value of $I_2$ 
as function of $\xn$ using the continuous limit $l/t\rightarrow u$
\bb\nn
I_2\simeq \frac{t^2}{\xn^2}\left [1+(\xn-1)\e^{\xn}\right],
\ee
and therefore, assuming that $t\gg 1$
\bb
\phi''=-t^2\left  
[\frac{1}{1-\e^{-\xn}}-\frac{1}{\xn}-\frac{1}{n}\right ]
\ee
This second derivative is negative only for $\xn>0$ and $n\ge 2$, since the function
$(1-\e^{-\xn})^{-1}-\xn^{-1}$ is larger than $1/2$ when $\xn>0$ and less than 
$1/2$ when $\xn<0$. 
%
\section{\label{app1}}
%
The size distribution $P_n(x,t)$ \eref{eq_size_n} is a multiple integral which can
be solved by using, as before, the integral representation of the Kronecker delta function.
One has indeed
\bb\fl\fl\nn
P_{n}(x,t)=
\frac{1}{n+1}\sum_{i=0}^n
\int_{-\pi}^{\pi}\frac{\drm y}{2\pi}\e^{iyt}
\int_x^1\frac{\drm u_1 \e^{-iy}}{1-u_1\e^{-iy}}
\cdots \int_x^{u_{i-1}}
\frac{\drm u_{i}\e^{-iy}}{1-u_i\e^{-iy}}\frac{\e^{-iy}}{1-(u_{i}-x)\e^{-iy}}
\\ \nn\times
\int_0^{u_i-x}
\frac{\drm u_{i+2}\e^{-iy}}{1-u_{i+2}\e^{-iy}}
\cdots\int_0^{u_{n-1}}
\frac{\drm u_{n}\e^{-iy}}{1-u_n\e^{-iy}}
\ee
Using integrations by parts from variables $u_1$ to $u_{i-1}$, successively with 
the integrand
\bb
\frac{\e^{-iy}}{1-u\e^{-iy}}\rightarrow \log\left (
\frac{1-\e^{-iy}}{1-u\e^{-iy}}\right )=-F(u,1)
\ee
and symmetrization of variables $u_{i+2},\cdots,u_n$, 
we can rewrite the previous expression as
\bb
P_{n}(x,t)=
\frac{1}{n+1}\sum_{i=0}^n
\int_{-\pi}^{\pi}\frac{\drm y}{2\pi}\e^{iyt}
\int_x^1\drm u
\frac{F(u,1)^{i-1}}{(i-1)!}
\frac{\e^{-iy}}{1-u\e^{-iy}}\frac{\e^{-iy}}{1-(u-x)\e^{-iy}}
\\ \nn\times
\frac{1}{(n-i-1)!}
\left (\int_0^{u-x}
\frac{\drm v\e^{-iy}}{1-v\e^{-iy}}\right )^{n-i-1}
\ee
The cases $i=0$ and $i=n$ has to be taken separately, and one obtains, after summing
over $i=1,\cdots,n-1$, 
\bb\fl\fl\nn
P_{n}(x,t)=
\frac{1}{n+1}
\int_{-\pi}^{\pi}\frac{\drm y}{2\pi}\e^{iyt}
\left \{
\frac{\e^{-iy}}{1-(1-x)\e^{-iy}}\frac{1}{(n-1)!}
\left (\int_0^{1-x}\frac{\drm u\e^{-iy}}{1-u\e^{-iy}}\right )^{n-1}
\right .
\\ \fl\fl \nn
\left .
+\frac{1}{(n-2)!}
\int_x^1\drm u
\frac{\e^{-iy}}{1-u\e^{-iy}}\frac{\e^{-iy}}{1-(u-x)\e^{-iy}}
\left [
-\log \left (\frac{[1-\e^{-iy}][1-(u-x)\e^{-iy}]}{1-u\e^{-iy}}\right ) \right ]^{n-2}
\right .
\\ \fl
\left .
+\frac{\e^{-iy}}{1-x\e^{-iy}}\frac{1}{(n-1)!}
\left (\int_x^{1}\frac{\drm u\e^{-iy}}{1-u\e^{-iy}}\right )^{n-1}
\right \}
\ee
The first term into bracket corresponds to $i=0$, the second term is the summation
over $i=1,\cdots,n-1$, and the last term corresponds to $i=n$. The second integral inside
the bracket can be performed exactly, and combined with the other two terms
\bb
\fl\fl\nn
P_{n}(x,t)=\frac{1}{x}
\frac{1}{n+1}
\int_{-\pi}^{\pi}\frac{\drm y}{2\pi}\e^{iyt}
\left \{
\frac{\e^{-iy}-1}{1-(1-x)\e^{-iy}}
\frac{\left [-\log (1-(1-x)\e^{-iy})\right ]^{n-1}}{(n-1)!}
\right .
\\ \fl\fl \nn
\left .
+\frac{1}{1-x\e^{-iy}}\frac{1}{(n-1)!}
\left [-\log\left (\frac{1-\e^{-iy}}{1-x\e^{-iy}}\right )\right ]^{n-1}
\right \}
\ee
This expression can be rewritten using a derivative with respect to $x$
\bb
\fl\fl\nn
P_{n}(x,t)=\frac{1}{(n+1)!}\frac{1}{x}\frac{\partial}{\partial x}
\int_{-\pi}^{\pi}\frac{\drm y}{2\pi}\e^{iyt}
\left \{
(\e^{iy}-1)\left [-\log (1-(1-x)\e^{-iy})\right ]^{n}
\right .
\\ \fl\fl 
\left .
-\e^{iy}
\left [-\log\left (\frac{1-\e^{-iy}}{1-x\e^{-iy}}\right )\right ]^{n}
\right \}
\ee
The logarithmic functions can be decomposed using \eref{def_stirling}, and the integration over
$y$ performed. After some algebra, one finds
\bb\label{Pn}
\fl\fl
P_{n}(x,t)=\frac{1}{(n+1)}\frac{1}{x}\frac{\partial}{\partial x}
\left [
-(1-x)^t\rho_n(t)-\sum_{r=n}^t\bin{t}{r-1}x^{t+1-r}(1-x)^r\rho_n(r)
\right ]
\ee
%
%
\section{\label{app2}}
%
One starts with the initial condition at time $s$, with the configuration $(v_1,\cdots,v_m)$ \eref{init_condQ}. 
We will note as in \eref{def_Qts}
\bb
Q_{t,s}^{(n)}(u_1,\cdots,u_n)=\langle u_1,\cdots,u_n;t|v_1,\cdots,v_m;s\rangle
\ee
by assuming implicitly the condition $(v_1,\cdots,v_m)$ at time $s$ to avoid cumbersome notations.
Since there is no obvious Green function in the problem, it is easier to deduce the generic form of this amplitude after few time steps. 
For example, at time $s+1$ one finds the set of solutions
\bb\nn
\left \{
\begin{array}{l}
Q_{s+1,s}^{(m+1)}(u_1,\cdots,u_{m+1})=\Delta_m,
\\ \nn
Q_{s+1,s}^{(m)}(u_1,\cdots,u_{m})=\Delta_{m-1},
\\ \nn
Q_{s+1,s}^{(m-1)}(u_1,\cdots,u_{m-1})=\Delta_{m-2},
\\ \nn\cdots
\\
Q_{s+1,s}^{(1)}(u_1)=\Delta_{0}
\end{array}
\right .
\ee
with the implicit condition $u_0=1>u_1>u_2>\cdots>u_m>u_{m+1}>\cdots$ at every time.
Here $\Delta_m=\prod_{i=1}^m\delta(u_i-v_i)$, $\Delta_{m-1}=\prod_{i=1}^{m-1}\delta(u_i-v_i)\theta(u_m-v_m)$, and
in general
\bb
\Delta_{m-k}=\prod_{i=1}^{m-k}\delta(u_i-v_i)\theta(u_{m-k+1}-v_{m-k+1})
\ee
The $\theta$ function is necessary since the last $u_i$ is either bounded by $u_{i-1}>u_{i}$ or by $v_{i-1}>u_i>v_i$. At time $s+2$, the solutions are given by
\begin{equation}
\left \{
\begin{array}{l}
Q_{s+2,s}^{(m+2)}(u_1,\cdots,u_{m+2})=\Delta_m,
\\ \label{eqs2}
Q_{s+2,s}^{(m+1)}(u_1,\cdots,u_{m+1})=\Delta_{m-1}+\Delta_mu_{m+1},
\\ \nn
Q_{s+2,s}^{(m)}(u_1,\cdots,u_{m})=\Delta_{m-2}+\Delta_{m-1}u_m,
\\ \nn\cdots
\\
Q_{s+2,s}^{(1)}(u_1)=\Delta_0u_1
\end{array}
\right .
\end{equation}
And for time $s+3$, one has
\begin{equation}\label{eqs3}
\left \{
\begin{array}{l}
Q_{s+3,s}^{(m+3)}(u_1,\cdots,u_{m+3})=\Delta_m
\\ \nn
Q_{s+3,s}^{(m+2)}(u_1,\cdots,u_{m+2})=\Delta_{m-1}+\Delta_m(u_{m+1}+u_{m+2}),
\\ \nn
Q_{s+3,s}^{(m+1)}(u_1,\cdots,u_{m+1})=\Delta_{m-2}+\Delta_{m-1}(u_m+u_{m+1})+\Delta_mu_{m+1}^2,
\\ \nn
Q_{s+3,s}^{(m)}(u_1,\cdots,u_{m})=\Delta_{m-3}+\Delta_{m-2}(u_{m-1}+u_{m})+\Delta_{m-1}u_{m}^2,
\\ \nn\cdots
\\
Q_{s,s+3}^{(1)}(u_1)=\Delta_0u_1^2
\end{array}
\right .
\end{equation}
From these equations, one can infer that at time $s+l=t$, if $m>l$
\bb\nn
Q_{s+l,s}^{(m)}(u_1,\cdots,u_m)=\Delta_{m-l}+\Delta_{m-l+1}\sum_{k_0+\cdots+k_{l-2}=1}u_{m-l+2}^{k_{l-2}}\cdots u_{m-1}^{k_{1}}u_{m}^{k_0}
\\
+\cdots+\Delta_{m-2}\sum_{k_0+k_1=l-2}u_{m-1}^{k_1}u_{m}^{k_0}+\Delta_{m-1}u_m^{l-1}
\ee
and if $m\le l$
\bb\nn
Q_{s+l,s}^{(m)}(u_1,\cdots,u_m)=\Delta_{0}\sum_{k_0+\cdots +k_{m-1}=l-m}u_{1}^{k_{m-1}}\cdots u_{m-1}^{k_{1}}u_{m}^{k_0}
\\
+\cdots+\Delta_{m-2}\sum_{k_0+k_1=l-2}u_{m-1}^{k_1}u_{m}^{k_0}+\Delta_{m-1}u_m^{l-1}
\ee
The different integrations over variables $u_i$s in \eref{Ctsd_def} can be performed iteratively. Taking the last three terms of the previous expressions, one has, after integration
\bb\nn
(a)\; \int\drm u \Delta_{m-1}u_m^{l-1}=\int_{v_m}^{v_{m-1}}u_m^{l-1}\drm u_m=\frac{1}{l}
\left (v_{m-1}^l-v_m^{l} \right ),
\\ \nn
(b) \int\drm u \Delta_{m-2}\sum_{k_0+k_1=l-2}u_{m-1}^{k_1}u_{m}^{k_0}
=\int_{v_{m-1}}^{v_{m-2}}\drm u_{m-1}
\int_{0}^{u_{m-1}}\drm u_m 
\sum_{k_0+k_1=l-2}u_{m-1}^{k_1}u_{m}^{k_0}
\\ \nn
=\frac{1}{l}
\left (v_{m-2}^l-v_{m-1}^{l} \right )\sum_{k_0+k_1=l-2}\frac{1}{k_0+1},
\\ \nn
(c) \int\drm u \Delta_{m-3}\sum_{k_0+k_1+k_2=l-3}u_{m-2}^{k_2}u_{m-1}^{k_1}u_{m}^{k_0}
=
\\ \nn
\int_{v_{m-2}}^{v_{m-3}}\drm u_{m-2}\int_{0}^{u_{m-2}}\drm u_{m-1}
\int_{0}^{u_{m-1}}\drm u_m 
\sum_{k_0+k_1+k_2=l-3}u_{m-2}^{k_2}u_{m-1}^{k_1}u_{m}^{k_0}
\\ 
=\frac{1}{l}
\left (v_{m-3}^l-v_{m-2}^{l} \right )\sum_{k_0+k_1+k_2=l-3}\frac{1}{k_0+1}\frac{1}{k_0+k_1+2}
\ee
Let us define the coefficients
\bb\nn
a_0(l)=1,\;a_1(l)=\sum_{k=1}^{l-1}\frac{1}{k},
\\ \label{ar_def}
a_r(l)=\sum_{l_0+\cdots+l_{r}=l,l_i\ge 1}\frac{1}{l_0}\frac{1}{l_0+l_1}\cdots\frac{1}{l_0+\cdots+l_{r-1}}
=
\frac{1}{r!}\sum_{l_0+\cdots+l_{r}=l,l_i\ge 1}\frac{1}{l_0}\frac{1}{l_1}\cdots\frac{1}{l_{r-1}}
\\ \nn
a_{l-1}(l)=\frac{1}{(l-1)!},\;a_l(l)=0
\ee
These coefficients satisfy the recursion relation
\bb\label{ar_rec}
a_r(l+1)-a_r(l)=\frac{1}{l}a_{r-1}(l),\;
a_r(l)=\sum_{k=r}^{l-1}\frac{a_{r-1}(k)}{k}
\ee
If we take the second definition of \eref{ar_def}, one can use the Kronecker delta integral representation to rewrite the sum
as 
\bb\nn
a_r(l)=\frac{1}{r!}\int_{-\pi}^{\pi}\frac{\drm y}{2\pi}\e^{iyl}
\left (\sum_{k=1}^{\infty}\frac{\e^{-iyk}}{k} \right )^r
\left (\sum_{k=1}^{l}\e^{-iyk} \right )
\\
=\frac{1}{r!}\int_{-\pi}^{\pi}\frac{\drm y}{2\pi}\e^{iyl}
\left [-\log(1-\e^{-iy})\right ]^r
\left (\sum_{k=1}^{l}\e^{-iyk} \right )
\ee
Using the series expansion of the logarithm \eref{def_stirling}, it is easy to express coefficients $a_r(l)$ 
as function of the densities
\bb\label{ar}
a_r(l)=\sum_{p=r}^{l-1}\rho_r(p)=l\rho_{r+1}(l)
\ee
The last identity comes from a recurrence relation between Stirling numbers \cite{book:NIST}.
The resulting integral $\int\drm u Q_{s+l,s}^{(m)}(u_1,\cdots,u_m)$ over variables $u_i$ is a polynom $\pol_l(v_0,v_1,\cdots,v_m)$ of the variables $v_i$  from the initial conditions at $t=s$ and implicitly present in the coefficients 
$\Delta$. This polynom (with $v_0=1$) can be written as
\bb\label{m_sup_l}\fl\fl
\pol_l(v_0,v_1,\cdots,v_m)&=&
\frac{1}{l}\left \{\sum_{r=1}^{l}v_{m-r}^l\left [ a_{r-1}(l)-a_{r}(l) \right ]
-v_m^la_0(l)\right \},\;\textrm{if $m\ge l$}
\\ \label{m_inf_l}
&=&\frac{1}{l}\left \{\sum_{r=1}^{m}v_{m-r}^l\left [ a_{r-1}(l)-a_{r}(l) \right ]
-v_m^la_0(l)+v_0^la_m(l)\right \},\;\textrm{if $m\le l$}
\ee
which is simply a sum of terms $v_i^l$. 
In the first case \eref{m_sup_l}, $\pol_l$ depends on $(v_{m-l},\cdots,v_{m})$ only whereas in 
the second case \eref{m_inf_l}, all variables $(v_0,\cdots,v_{m})$ are involved. Since we consider
$l=t-s\ge s\ge m$, the latter case is relevant for computing all coefficients in \eref{Ctsd_def}.
For example, when $l=4$ and $m=3$ one has
\bb
\pol_{4}(v_0,v_1,v_2,v_3)=\frac{1}{4}v_0^4+\frac{5}{24}v_1^4-\frac{5}{24}v_2^4-\frac{1}{4}v_3^4
\ee
%
\section{\label{app3}}
%
In this appendix, we perform the average of the polynom $\pol_l(v_1,\cdots,v_m)$, defined in the previous
appendix, over
all the possible configurations $(v_1,\cdots,v_m)$ weighted by $Q_s^{(m)}(v_1,\cdots,v_m)$ as defined
in the disconnected part of the correlation function \eref{Ctsd_def}. This is tantamount to perform the multiple integrals
\bb\fl\nn
C_{0,m}(l)=\rho_m(s),
\\ \fl\label{Ckm_def}
C_{k,m}(l)=\int_0^1\drm v_1\int_0^{v_1}\drm v_2\cdots\int_0^{v_{m-1}}\drm v_m
\left ( \sum_{k_1+\cdots+k_m=s-m}v_1^{k_1}v_2^{k_2}\cdots v_m^{k_m} \right )v_k^l
\ee
for each term $v_k^l$ of the polynom $\pol_l$. We can then express $C_d(t,s)$ as function of coefficients 
$a_r(l)$ and $C_{k,m}(l)$ only as seen further below. Using the Kronecker delta integral representation 
\bb
C_{k,m}(l)=\int\drm v \int_{-\pi}^{\pi}\frac{\drm y}{2\pi}\e^{iy(s-m)}\frac{1}{1-v_1\e^{-iy}}
\cdots \frac{v_k^l}{1-v_k\e^{-iy}}\cdots \frac{1}{1-v_m\e^{-iy}}
\ee
with $C_{0,m}(l)=\rho_m(s)$, we can perform successive integrations by parts, and the previous multiple integral can be reduced to a more simple form which leads to
\bb\label{Ckm}
C_{k,m}(l)=\int_{-\pi}^{\pi}\frac{\drm y}{2\pi}\frac{\e^{iy(s-m)}}{(k-1)!(m-k)!}
\int_0^1\drm vF(v,1)^{k-1}\frac{v^l}{1-v\e^{-iy}}F(0,v)^{m-k}
\ee
where $F(v,1)=-\e^{iy}\log\left [(1-\e^{-iy})/(1-v\e^{-iy})\right ]$ and 
$F(0,v)=-\e^{iy}\log(1-v\e^{-iy})$. Here the integers are restricted to the interval 
$1\le k\le m\le s$. Using \eref{def_stirling} we expand $F(v,1)^{k-1}$ and $F(0,v)^{m-k}$ in order 
to express $C_{k,m}$ as products of two density functions
\bb\fl\nn
C_{k,m}(l)=
\sum_{r\ge k-1}\sum_{r'\ge m-k}\rho_{k-1}(r)\rho_{m-k}(r')
\int_{-\pi}^{\pi}\frac{\drm y}{2\pi}\e^{iy(s-1-r-r')}\int_0^1\drm v\frac{(1-v)^rv^{l+r'}}{(1-v\e^{-iy})^{r+1}}
\\ \label{Ckm_sol}
=\sum_{r=k-1}^{t-2}\sum_{r'=m-k}^{s-2-r}\rho_{k-1}(r)\rho_{m-k}(r')\frac{(s-2-r')!(t-2-r)!}{(s-2-r'-r)!(t-1)!}
\ee
This is an exact expression. Using \eref{ar}, \eref{Ckm_sol}, and 
$C_{0,m}(l)=\rho_m(s)$, one can compute the configuration average of the
polynomials  $\pol_l(v_0,v_1,\cdots,v_m)$ for every $m$ and expand the 
auto-correlation function as a sum over decreasing terms for $l>s$, since the 
configuration average of $v_k^l$ decreases with increasing $k$ for large $l$, 
the dominant term being $v_0^l=1$
\bb
C_d(t,s)=\sum_{m=1}^s\left \{
\rho_m(s)\rho_m(t-s)+C_{1,m}[\rho_{m-1}(t-s)-\rho_m(t-s)]+\cdots
\right \}
\ee
Coefficient $C_{1,m}$ can be computed exactly, using $\rho_0(r)=\delta_{r,0}$ and the following recurrence identity \cite{book:NIST}
\bb
\rho_m(s)=\frac{1}{s}\sum_{r=m-1}^{s-1}\rho_{m-1}(r)
\ee
which leads to the simple result
\bb
C_{1,m}=\frac{s-1}{t-1}\rho_m(s-1)
\ee
Therefore $C_d(t,s)$ can be expanded as \eref{Ctsd_sol}. When $t\gg s\gg 1$, 
coefficients $C_{km}$ in \eref{Ckm_sol} are of the order of $(s/t)^{k}$, or
\bb
C_{k,m}\simeq \left (\frac{s}{t}\right )^k\frac{\rho_{m-k+1}(s-k)}{(k-1)!}
\ee
after assuming that only the term $r=k-1$ in the sum gives the main 
contribution at this order. This provides a controlled expansion in $(s/t)$ for 
the auto-correlation function.

\section*{References}
\bibliography{biblio_frag}

\providecommand{\newblock}{}
\begin{thebibliography}{10}
\expandafter\ifx\csname url\endcsname\relax
  \def\url#1{{\tt #1}}\fi
\expandafter\ifx\csname urlprefix\endcsname\relax\def\urlprefix{URL }\fi
\providecommand{\eprint}[2][]{\url{#2}}

\bibitem{struick77}
Struik L~C~E 1977 {\em Polymer Engineering \& Science\/} {\bf 17} 165--173

\bibitem{book:henkel_aging}
Malte~Henkel Michel~Pleimling R~S 2007 {\em Ageing and the Glass Transition\/}
  1st ed Lecture notes in physics 716 (Springer) section 1.3.1: scaling forms

\bibitem{book:henkel_noneq}
Malte~Henkel Haye~Hinrichsen S~L 2008 {\em Non-Equilibrium Phase Transitions:
  Absorbing Phase Transitions\/} 1st ed ({\em Theoretical and Mathematical
  Physics\/} vol volume 1) (Springer)

\bibitem{crisanti03}
Crisanti A and Ritort F 2003 {\em J. Phys. A: Math. Gen.\/} {\bf 36} R181--R290

\bibitem{book:Struick}
Struick L~C~E 1978 {\em Physical ageing in amorphous polymers and other
  materials\/} (Amsterdam: Elsevier)

\bibitem{krapivsky00}
Krapivsky P~L, Grosse I and Ben-Naim E 2000 {\em Phys. Rev. E\/} {\bf 61} R993

\bibitem{book:turcotte}
Turcotte D~L 1997 {\em Fractals and Chaos in Geology and Geophysics\/} 2nd ed
  (Cambridge University Press)

\bibitem{brown89}
Brown W~K 1989 {\em J. Astrophys. Astr.\/} {\bf 10} 89--112

\bibitem{herrmann06}
Herrmann H~J, Wittel F~K and Kun F 2006 {\em Physica A: Statistical Mechanics
  and its Applications\/} {\bf 371} 59 -- 66

\bibitem{bershadskii00}
Bershadskii A 2000 {\em J. Phys. A: Math. Gen.\/} {\bf 33} 2179--2183

\bibitem{ziff:86}
Ziff R~M and McGrady E~D 1986 {\em Macromolecules\/} {\bf 19} 2513--2519

\bibitem{hassan:95}
Hassan M and Rodgers G 1995 {\em Phys. Lett. A\/}  95--98

\bibitem{jyf:2013}
Fortin J~Y, Mantelli S and Choi M 2013 {\em Journal of Physics A: Mathematical
  and Theoretical\/} {\bf 46} 225002

\bibitem{book:krapivsky}
Krapivsky P~L, Redner S and Ben-Naim E 2010 {\em A Kinetic View of Statistical
  Physics\/} (Cambridge University Press)

\bibitem{book:bertoin}
Bertoin J 2006 {\em Random Fragmentation and Coagulation Processes\/} 1st ed
  Cambridge Studies in Advanced Mathematics (Cambridge University Press)

\bibitem{stemmer94}
Stemmer W~P 1994 {\em Proceedings of the National Academy of Sciences\/} {\bf
  91} 10747--10751

\bibitem{cohen10}
Cohen J 2001 {\em Science\/} {\bf 293} 237--237

\bibitem{lobo12}
Lobo D, Beane W~S and Levin M 2012 {\em PLOS Computational Biology\/} {\bf 8}
  1--12

\bibitem{azais15}
Azaïs R and Genadot A 2015 {\em TEST\/} {\bf 24} 341--360

\bibitem{ke03}
Ke J, Lin Z and Zhuang Y 2003 {\em Eur. Phys. J. B\/} {\bf 36} 423--428

\bibitem{ke04}
Ke J, Cai X~O and Lin Z 2004 {\em Physics Letters A\/} {\bf 331} 281 -- 287

\bibitem{robe:16}
Robe D~M, Boettcher S, Sibani P and Yunker P 2016 {\em EPL (Europhysics
  Letters)\/} {\bf 116} 38003

\bibitem{yunker:09}
Yunker P, Zhang Z, Aptowicz K~B and Yodh A~G 2009 {\em Phys. Rev. Lett.\/} {\bf
  103}(11) 115701

\bibitem{sibani:93}
Sibani P and Littlewood P~B 1993 {\em Phys. Rev. Lett.\/} {\bf 71}(10)
  1482--1485

\bibitem{sibani:03}
Sibani P and Dall J 2003 {\em EPL (Europhysics Letters)\/} {\bf 64} 8

\bibitem{sibani:16}
Sibani P and Boettcher S 2016 {\em Phys. Rev. E\/} {\bf 93}(6) 062141

\bibitem{ritort:95}
Ritort F 1995 {\em Phys. Rev. Lett.\/} {\bf 75} 1190

\bibitem{franz:96}
Franz S and Ritort F 1996 {\em Journal of Stat. Phys.\/} {\bf 85} 131--150

\bibitem{megen:94}
van Megen W and Underwood S~M 1994 {\em Phys. Rev. E\/} {\bf 49}(5) 4206--4220

\bibitem{jackle:91}
J{\"a}ckle J and Eisinger S 1991 {\em Zeitschrift f{\"u}r Physik B Condensed
  Matter\/} {\bf 84} 115--124

\bibitem{faggionato:12}
Faggionato A, Martinelli F, Roberto C and Toninelli C 2013  {\bf 19} 407--452
  (\textit{Preprint} \eprint{arXiv:1205.1607})

\bibitem{Abram_Stirl}
Danos M 1984 {\em Pocketbook of Mathematical Functions - Abramowitz and Stegun
  abbreviated\/} (H. Deutsch) page 367, Stirling Numbers of the First Kind,
  26.1.3

\bibitem{Gould}
Jocelyn~Quaintance H~W~G 2015 {\em Combinatorial Identities for Stirling
  Numbers: The Unpublished Notes of H W Gould\/} (World Scientific Publishing
  Company)

\bibitem{Delannay:1996}
Delannay R, Ca\"er G~L and Botet R 1996 {\em Journal of Physics A: Mathematical
  and General\/} {\bf 29} 6693

\bibitem{Luck:2008}
Godr\`eche C and Luck J~M 2008 {\em Journal of Statistical Mechanics: Theory
  and Experiment\/} {\bf 2008} P11006

\bibitem{BenNaim:2002}
Ben-Naim E and Krapivsky P~L 2002 {\em Journal of Physics A: Mathematical and
  General\/} {\bf 35} L557

\bibitem{BenNaim:2004}
Ben-Naim E and Krapivsky P~L 2004 {\em Journal of Physics A: Mathematical and
  General\/} {\bf 37} 5949

\bibitem{Wilf:93}
Wilf H~S 1993 {\em Journal of Combinatorial Theory\/} {\bf 64} 344--349

\bibitem{Temme:93}
Temme N 1993 {\em Studies in Applied Mathematics\/} {\bf 89} 233--243

\bibitem{Durang:2014}
Durang X, Henkel M and Park H 2014 {\em Journal of Physics A: Mathematical and
  Theoretical\/} {\bf 47} 045002

\bibitem{Durang:2011}
Durang X, Fortin J~Y and Henkel M 2011 {\em Journal of Statistical Mechanics:
  Theory and Experiment\/} {\bf 2011} P02030

\bibitem{book:NIST}
Olver F~W~J, Lozier D~W, Boisvert R~F and Clark C~W 2010 {\em NIST handbook of
  mathematical functions\/} 1st ed (Cambridge University Press) recurrence
  relation 26.8.20

\end{thebibliography}

\end{document}